\documentclass[prb,aps,twocolumn,superscriptaddress,showpacs]{revtex4-1}

\usepackage{graphicx}
\usepackage{epstopdf}
\usepackage[colorlinks=true,linkcolor=black, citecolor=blue, urlcolor=blue, 
    unicode=true]{hyperref}

\begin{document}

\title{Molecular orientational melting within a lead-halide octahedra framework - the order-disorder transition in CH$_{3}$NH$_{3}$PbBr$_{3}$}

\author{K. L. Brown}
\affiliation{School of Physics and Astronomy, University of Edinburgh, Edinburgh EH9 3JZ, UK}

\author{S. F. Parker}
\affiliation{ISIS Facility, Rutherford Appleton Laboratory, Chilton, Didcot OX11 0QX, UK}

\author{I. Robles Garc\'ia}
\affiliation{School of Physics and Astronomy, University of Edinburgh, Edinburgh EH9 3JZ, UK}

\author{S. Mukhopadhyay}
\affiliation{ISIS Facility, Rutherford Appleton Laboratory, Chilton, Didcot OX11 0QX, UK}
\affiliation{Department of Materials, Imperial College London, Exhibition Road, London SW7 2AZ, UK}

\author{V. Garc\'ia Sakai}
\affiliation{ISIS Facility, Rutherford Appleton Laboratory, Chilton, Didcot OX11 0QX, UK}

\author{C. Stock}
\affiliation{School of Physics and Astronomy, University of Edinburgh, Edinburgh EH9 3JZ, UK}

\date{\today}

\begin{abstract}
Lead-halide organic-inorganic perovskites consist of an inorganic host framework with an organic molecule occupying the interstitial space.  The structure and dynamics of these materials have been heavily studied recently due to interest in their exceptional photovoltaic properties.  We combine inelastic neutron scattering, Raman spectroscopy, and quasielastic neutron scattering to study the temperature dependent dynamics of the molecular cation in CH$_{3}$NH$_{3}$PbBr$_{3}$.   By applying high resolution quasielastic neutron scattering, we confirm the [CH$_{3}$NH$_{3}$]$^{+}$ ions are static in the low temperature orthorhombic phase yet become dynamic above 150 K where a series of structural transitions occur.  This molecular melting is accompanied by a temporal broadening in the intra-molecular modes probed through high energy inelastic spectroscopy.  Simultaneous Raman measurements, a strictly $|Q|$=0 probe, are suggestive that this broadening is due to local variations in the crystal field environment around the hydrogen atoms. These results confirm the strong role of hydrogen bonding and also a coupling between molecular and framework dynamics.

\end{abstract}

\maketitle

\section{Introduction}

Organic-inorganic perovskites have recently received a lot of attention due to their potential applications in optoelectronics~\cite{Saparov}. These compounds are of the perovskite structure ABX$_{3}$, where in this case A is an organic cation (typically methylammonium (MA: [CH$_{3}$NH$_{3}$]$^{+}$) or formamidinum (FA: [HC(NH$_{2}$)$_{2}$]$^{+}$), though others have been investigated), B is a metal (Pb or Sn), and X is a halogen (Cl, Br or I). In particular, MAPbI$_{3}$ has been investigated in depth after it was found to have desirable photovoltaic properties, with thin-film solar panels made of this material or MAPb(I$_{1-x}$Br$_{x}$)$_{3}$ achieving efficiencies of above 20\%\cite{Park, Noh, Zhou, Herz, Bennett}.  The key differences between the three different halide ions in MAPbX$_{3}$ (X = I, Br, Cl) is their nuclear radii. This radius  affects the size of the inorganic cages in which the MA cation sits,  and thus the exact radius of rotation accessible for the molecule. In  the case of Cl, the decreased size of the halide ion has been linked  to the presence of two potential orientational states of the cation in  the low temperature phase, as opposed to the single orientation found  when I or Br is present instead\cite{Poglitsch, Brock, Ren, Mashiyama,  Whitfield}. The choice of halide anion also has an impact on the exact optical properties of the system, and a mixture of anions can be used to tune the band gap of the system\cite{Park, Noh, Zhou, Herz}.

The behavior of the organic cation in these hybrid organic-inorganic compounds is recognized to be key to understanding their overall properties, including the photovoltaic properties and how this material can be implemented in solar panel devices\cite{Herz, Gao, Zhu}. Importantly, the molecule is able to rotate in the perovskite cage, and the exact nature of this rotation and how it depends on the structural phase of the system is still yet to be understood. As the temperature is lowered, MAPbBr$_{3}$ experiences multiple structural phase transitions, driven by the perovskite rotational modes\cite{Fuji, Brivio, Comin, Lee, Lee2, Letoubon, Swainson, Kawamura}. At room temperature MAPbBr$_{3}$ is cubic (phase I) undergoing a transition to a tetragonal unit cell (phase II) at 225 K.  At 155 K a second structural transition occurs into an unknown phase (phase III) which is fitted by a tetragonal space group, but believed by some to be incommensurate, with a final structural transition occurring at 150 K to an orthorhombic unit cell (phase IV). The structure of these phases was first solved by Poglitsch and Weber, who identified phase I as being Pm3m, $a$ = 5.90 \AA; phase II as being I4/mcm, $a$  = 8.32 \AA, $c$ = 11.83 \AA; phase III as being tetragonal P4/mmm, $a$=5.89 \AA, $c$=5.86 \AA; and phase IV as Pna2$_{1}$, $a$ = 7.97 \AA, $b$ = 8.58 \AA, $c$ = 11.85 \AA \space (Ref. \onlinecite{Poglitsch}). It has been suggested that the motion of the perovskite cages is strongly coupled with the behavior of the organic cation\cite{Lee, Lee2, Ong}. It should be noted that MAPbBr$_{3}$ is structurally identical to MAPbI$_{3}$ in the cubic phase, however, MAPbBr$_{3}$ is found to have different space groups in the tetragonal and orthorhombic phases (phases II and IV in MAPbBr$_{3}$), and displays an additional intermediate phase, which we refer to as phase III\cite{Kawamura, Poglitsch, Mashiyama2, Page, Swainson2}.

The organic-inorganic perovskites consist of a MA ion occupying the interstitial sites of a lead-halide framework.  The molecules' centre of mass gives positional order, however the orientational order depends on the local bonding and thermal fluctuations.  It is tempting to draw parallels between this and liquid crystals, specifically the melting of the smectic phase to the nematic phase, though in reality the two systems are not very similar\cite{Brock2}. In the nematic phase of liquid crystals there is an orientational order, but no positional order\cite{AlsNielsen, Pindak} whereas in the smectic phase both positional and orientational order is obtained~\cite{Brock, Ren, Mashiyama, Whitfield}. The situation in MAPbBr$_{3}$ is somewhat analogous to the molecular ionic crystals calcite and sodium nitrate where molecular positions are fixed by the crystalline lattice, however transitions from orientational order to disorder occurs with increasing temperatures\cite{Hagen, Harris}.  Such a transition has been termed an order-disorder transition and in calcite is characterized by a soft vertical ``column" of scattering at the zone boundary\cite{Hagen, Harris}; this contrasts with the soft mode which drives displacive transitions~\cite{Shirane}.  

Previous phonon work on MAPbBr$_{3}$ has identified a soft zone boundary acoustic phonon which softens at $\sim$~150 K and is accompanied by a temporally broad relaxational response at high temperatures~\cite{Swainson}.  Similar studies have been performed on MAPbI$_{3}$\cite{Leguy, Druzbiki}.  In this paper, we study the response of the molecular vibrations in MAPbBr$_{3}$ through high resolution quasielastic scattering and simultaneous Raman and inelastic neutron scattering measurements. We will show that the transition from the low temperature orthorhombic phase to higher temperature structures at 150 K is accompanied by an increase in the number of active, and freely rotating, molecular ions.  This is accompanied by an increase in the linewidth of the internal molecular motions.  This strong coupling between molecular motions and the PbBr$_{3}$ framework lattice indicates the importance of hydrogen bonding.

\section{Experimental}

A powder sample of MAPbBr$_{3}$ was prepared by the reaction of stoichiometric amounts of lead acetate and methylamine hydrobromide in hydrobromic acid, then evaporating away the excess acid to leave an orange colored precipitate. This precipitate was washed with diethyl ether to remove any residual acid in the sample.  For all measurements discussed below, the sample was wrapped in niobium foil to protect from reaction of any residual acid with the outer aluminum sample holders.  Care was taken to always store the sample in a dry atmosphere to prevent water absorption or possible degradation.  As discussed below, multiple experiments were performed on different samples with differing thermal histories providing the same results.

High resolution quasielastic measurements probing the low energy molecular fluctuations were carried out on IRIS\cite{Campbell} at ISIS (Rutherford Appleton Labs, UK), using PG002 analyzers with a fixed final energy of 1.843 meV providing a dynamic range of $\pm$ 0.5 meV.  An empty can background subtraction was carried out from a measurement performed at room temperature.  The resolution was measured from the base temperature response of the MAPbBr$_{3}$ sample, and was found to be 23.8~$\mu$eV from a Gaussian fit (full-width at half-maximum).  The IRIS spectrometer also features a diffraction detector bank at $2\theta \sim 170^{\circ}$, allowing for simultaneous diffraction data to be taken for a small range of {\bf Q} set by the time channel settings for the quasielastic measurements.  This afforded a measurement of the structural transitions simultaneously while measuring excitations.

High energy inelastic neutron scattering measurements to study the internal molecular motions were carried out on the MAPS spectrometer\cite{Perring} at ISIS (Rutherford Appleton Labs, UK), using the high resolution Fermi A chopper in parallel with a $t_{0}$ chopper spun at the proton source repetition rate of 50 Hz to eliminate high energy neutrons. Three incident energies were used for these measurements, with different associated Fermi chopper frequency: 60 meV with 100 Hz chopper; 250 meV with 400 Hz; and 650 meV with 600 Hz. The full-width half-maximum energy resolutions, at the elastic position, for these measurements were 2 meV, 6 meV and 16 meV respectively. Simultaneous to these spectroscopy measurements, Raman spectra were collected using a modified Renishaw InVia spectrometer\cite{Adams} with a wavelength of 785 nm. Two gratings of 1200 lines per mm and 1800 lines per mm were used, which provided measurements of different spectral ranges, giving a total range of 20 to 3200~cm$^{-1}$ (2.5 to 396.8~meV). 

\subsection{Quasielastic Neutron Scattering using IRIS- Molecular Motions}

\begin{figure}[b]
\includegraphics[width=9cm] {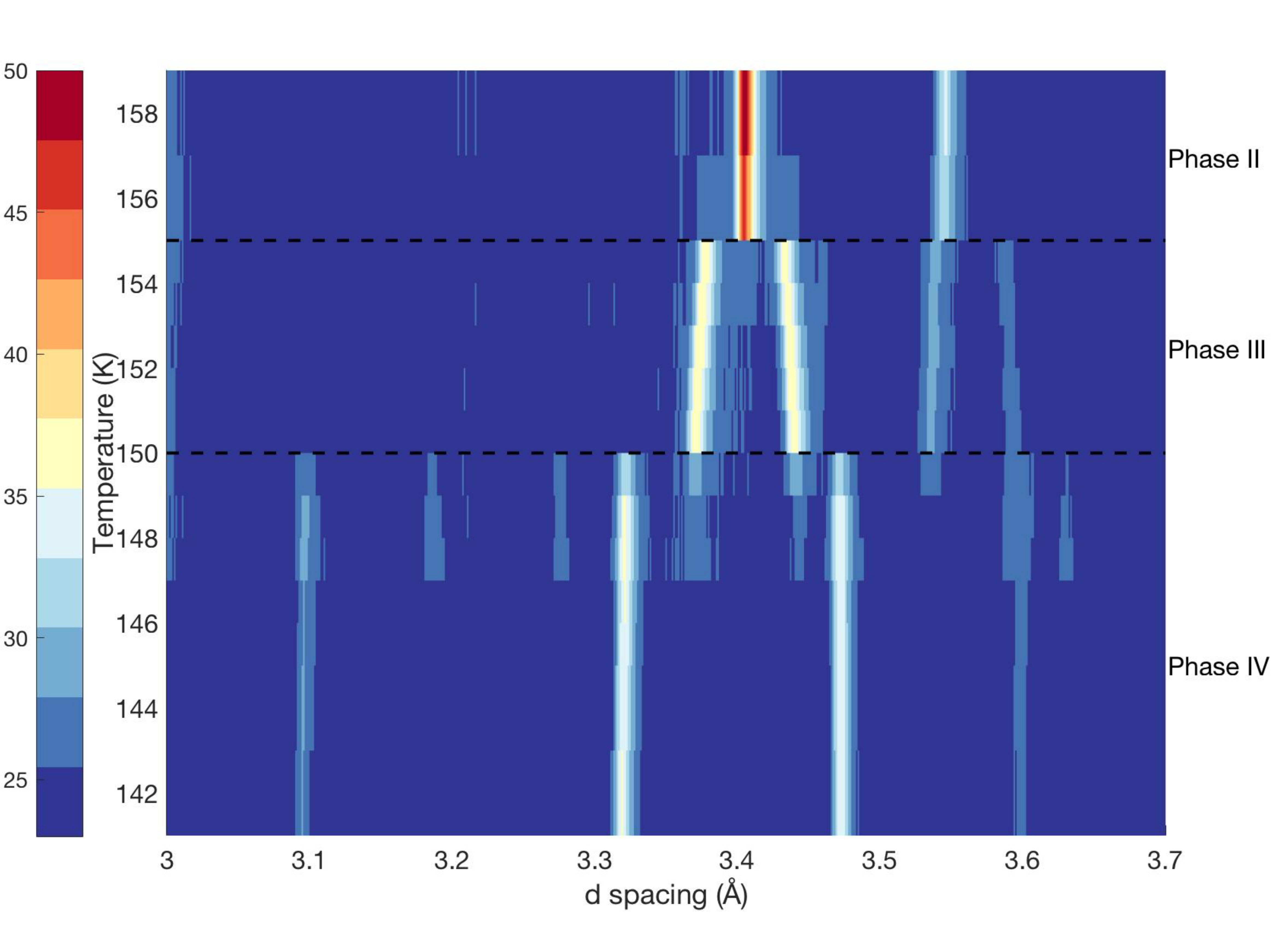}
\caption{\label{DiffractionData} The diffraction data obtained on IRIS for MAPbBr$_{3}$, showing the phase transitions II to III at 155 K and III to IV at 150 K.  The data was obtained simultaneously while performing quasielastic measurements probing the molecular motions.  The limited range in $d$ spacing is due to the constraints in time imposed by the inelastic measurements.}
\end{figure}

We first investigate the static structural properties of our sample using the diffraction detectors on IRIS.  MAPbBr$_{3}$ experiences four structural phase transitions between the cubic room temperature phase (phase I) and the tetragonal phase at base temperature (phase IV), with phase II being orthorhombic and phase III being previously identified as both incommensurate and tetragonal\cite{Poglitsch, Mashiyama2, Page, Swainson2}. The diffraction data obtained on IRIS, shown in Fig. \ref{DiffractionData}, can be used to confirm the presence of two phase transitions: the first at 150 K (IV to III), the second at 155 K (III to II). Some coexistence can be seen  between the phases, most noticeably just below 150 K around 3.4 \AA\ and 3.5 \AA. It should be noted that the extra lines seen only between 147 and 150 K not associated with phase II are present at lower temperatures, but counting times were not long enough to discern them clearly in the false color image. The range of {\bf Q} space available on IRIS is not large enough to substantiate a structural refinement to allow a comparison to the space groups found by Poglitsch and Weber\cite{Poglitsch}. These measurements confirm the structural transitions and the presence of an intermediate phase between the orthorhombic and tetragonal phases in our solution-synthesized samples.

\begin{figure}[t]
\includegraphics[width=9.5cm] {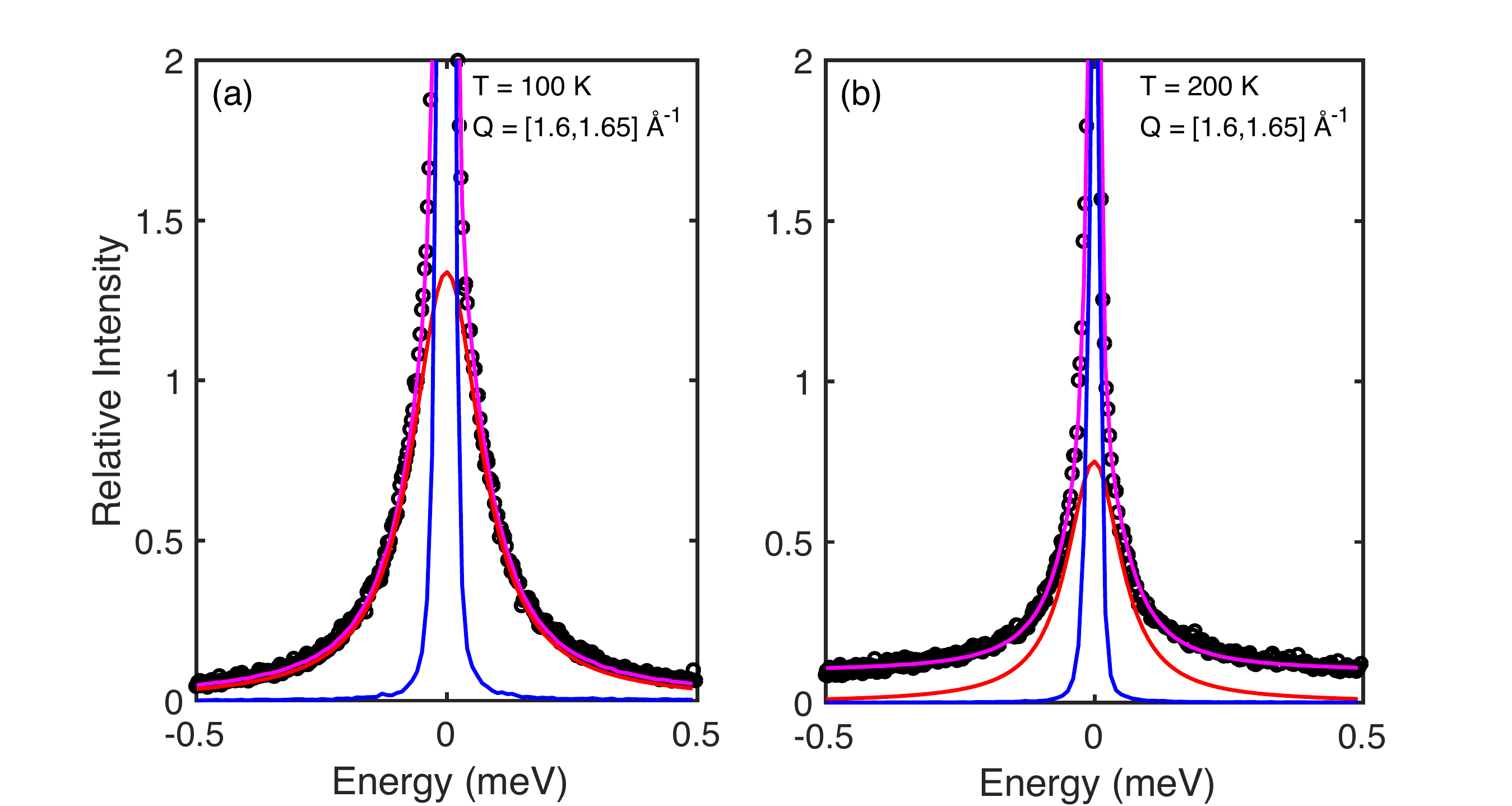}
\caption{\label{Lineshape1} Examples of the spectra obtained on IRIS at $(a)$ 100 K and $(b)$ 200 K. The black circles show the data points, the pink line shows the total fit, the blue shows the resolution delta function and the red shows the dynamical contribution. The additional increased baseline due to dynamical contributions outside the dynamic range is not explicitly plotted.}
\end{figure}

\begin{figure}[t]
\includegraphics[width=8cm] {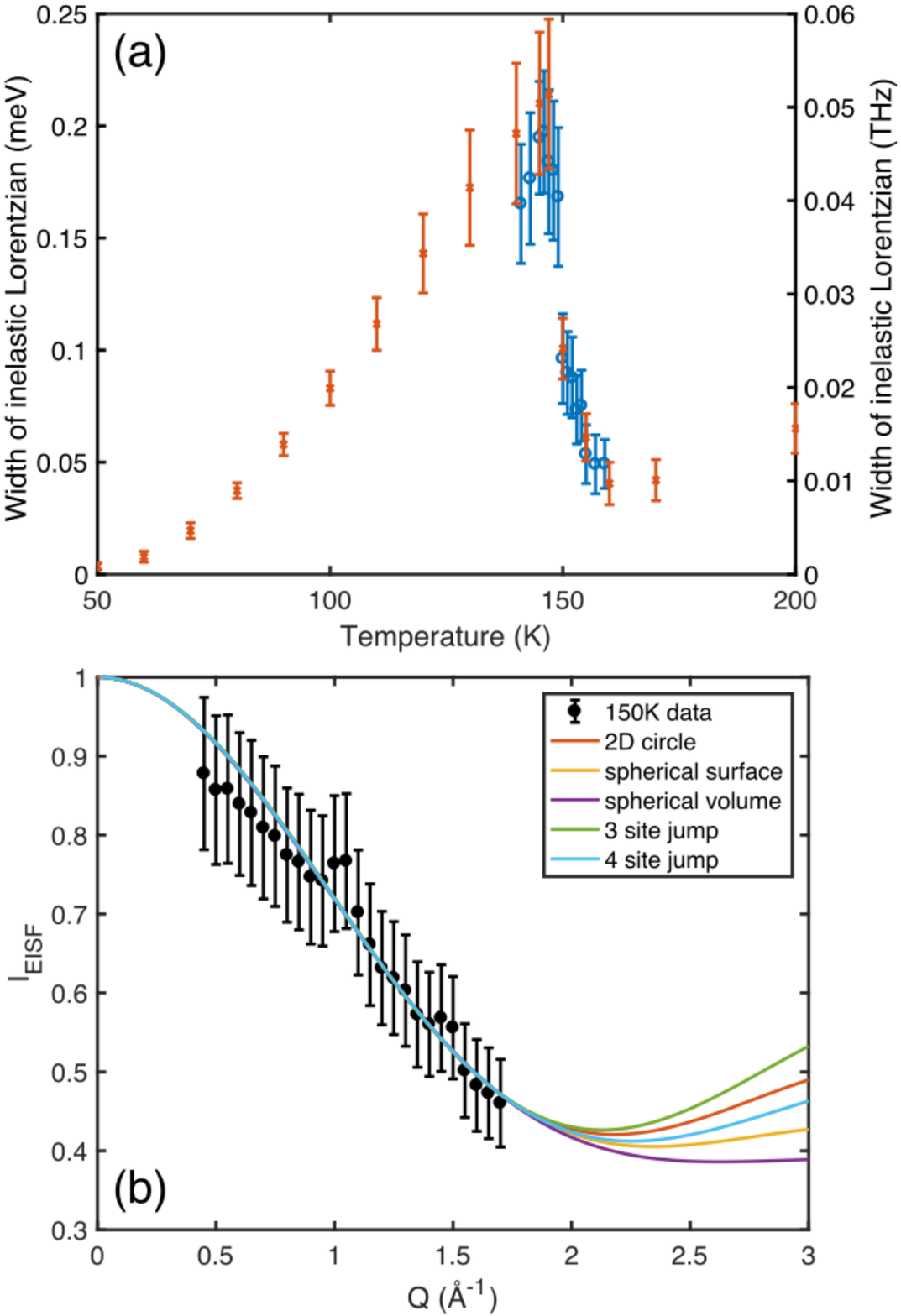}
\caption{\label{Lineshape2} $(a)$ The variation in width of the $|Q|$ averaged fitted quasielastic component with temperature. Blue and red data points are from two separate experiments. $(b)$ Trial fits to the $I_{EISF}(Q)$ at 150 K.  A systematic error for multiple scattering has been included in I$_{EISF}(Q)$.}
\end{figure}

Having confirmed the structural transitions, we now discuss the molecular dynamics.  Because the neutron incoherent cross section of H is over an order of magnitude larger than those of the non-H atoms, the neutron cross section measured is dominated by H dynamics~\cite{Sears}. Underlying this assumption is the fact that the incoherent cross section for H is 80.3 barns with the next largest incoherent scatter in MAPbBr$_{3}$ being N with a cross section of 0.5 barns.  We now use this and discuss the low energy quasielastic data sensitive to molecular motions based on data taken on the IRIS backscattering spectrometer.   

Two example spectra from IRIS are shown in Fig. \ref{Lineshape1} taken at 100 K and 200 K. The solid red curve is a fit to the sum of an elastic and dynamic component 

\begin{eqnarray}
S(E)=[I_{el}+I_{qe}]=I_{el} \delta(E) + \frac{I_{dyn}} {[1+(E/\gamma)^{2}]} \nonumber
\label{eq:1}
\end{eqnarray} 

\noindent where $2\gamma$ is the full-width at half-maximum of the dynamic component, convolved with the measured resolution function $\delta(E)$.  The linewidth 2$\gamma$ was found not to vary with momentum indicating no measurable diffusion.  This conclusion, and our analysis of the energy linewidth, is decoupled from the momentum dependence and therefore is independent of the model applied to the molecular motion and corresponding $Q$-dependence discussed below.  A plot of how this quasielastic full-width ($2\gamma$) averaged across the entire measured $|Q|$ range changes with temperature is shown in Fig. \ref{Lineshape2}$(a)$, showing an initial increase, indicative of shortened lifetimes, with increasing temperature followed by a distinct drop at the 150 K phase transition, which only begins to increase again after $\sim$160 K. Two separate experiments are plotted in different colors as the exact temperature of the phase transition is expected to be dependent on the thermal history\cite{Whitfield}; here we find the results from our separate experiments agree within error.   

It should be noted that, in the fits from which a full-width $2\gamma$ was obtained (examples of which are shown in Fig. \ref{Lineshape1}), for temperatures above 150 K a second faster component existed which was too broad to reliably fit over the $\pm$ 0.5 meV dynamic range, and was instead fitted as a constant increased baseline (see Fig. \ref{Lineshape1} $(b)$ for an example). This means that a second rotation of shorter timescale, which is not possible to measure with this instrument, is becoming activated at this point and has an energy scale outside the resolution of the IRIS spectrometer.  Indeed, work reported in Ref. \onlinecite{Swainson} suggests the presence of faster fluctuations above the 150 K transition with an energy scale on the order meV.   This is further corroborated by real-time vibrational spectroscopy~\cite{Bakulin} which identifies a fast motion around a particular axis and a slow ``jump-like" motion associated with a molecular dipole reorientation with similar timescales to that measured at high temperatures by NMR~\cite{Knop}.  Based on this comparison we conclude that the molecular motions are crossing over from ``wobbling" around a particular axis to full body motions as temperature is increased.    The assignment of the lower temperature response to a ``wobbling" of the molecular around a particular axis is also consistent with quasielastic neutron scattering data taken on the iodine variant~\cite{Leguy2}.

We now examine the elastic incoherent structure factor (EISF) 

\begin{eqnarray}
I_{EISF}(Q) = \frac{S_{el}(Q)} {S_{el}(Q) + S_{qe}(Q)} \nonumber
\label{eq:2}
\end{eqnarray} 

\noindent where $S_{el}(Q)$ and $S_{qe}(Q)$ are integrated over all energies. The momentum dependence of the EISF is sensitive to the real space nature of the equilibrium molecular motions. In our calculations of the EISF, the intensity of the elastic peak $S_{el}(Q)$ is taken as the area underneath the resolution convolved delta function component of the fit, and the sum $S_{el}(Q) + S_{qe}(Q)$ is taken as the area underneath the total fit. This means that despite the fact that the second, broader Lorentzian seen at higher temperatures is fitted as an increased baseline, it is still included in the calculations of the EISF. The measured $I_{EISF}(Q)$ was fitted to a function $A(Q)$ dependent on the proportion of rotators $p$ and radius of rotation $r$ as follows: 

\begin{eqnarray}
A(Q) = p + (1-p)f(Qr) \nonumber
\label{eq:2}
\end{eqnarray} 

\noindent where $f(Qr)$ is a model chosen for the nature of the molecular rotation. 

\begin{figure}[t]
\includegraphics[width=9cm] {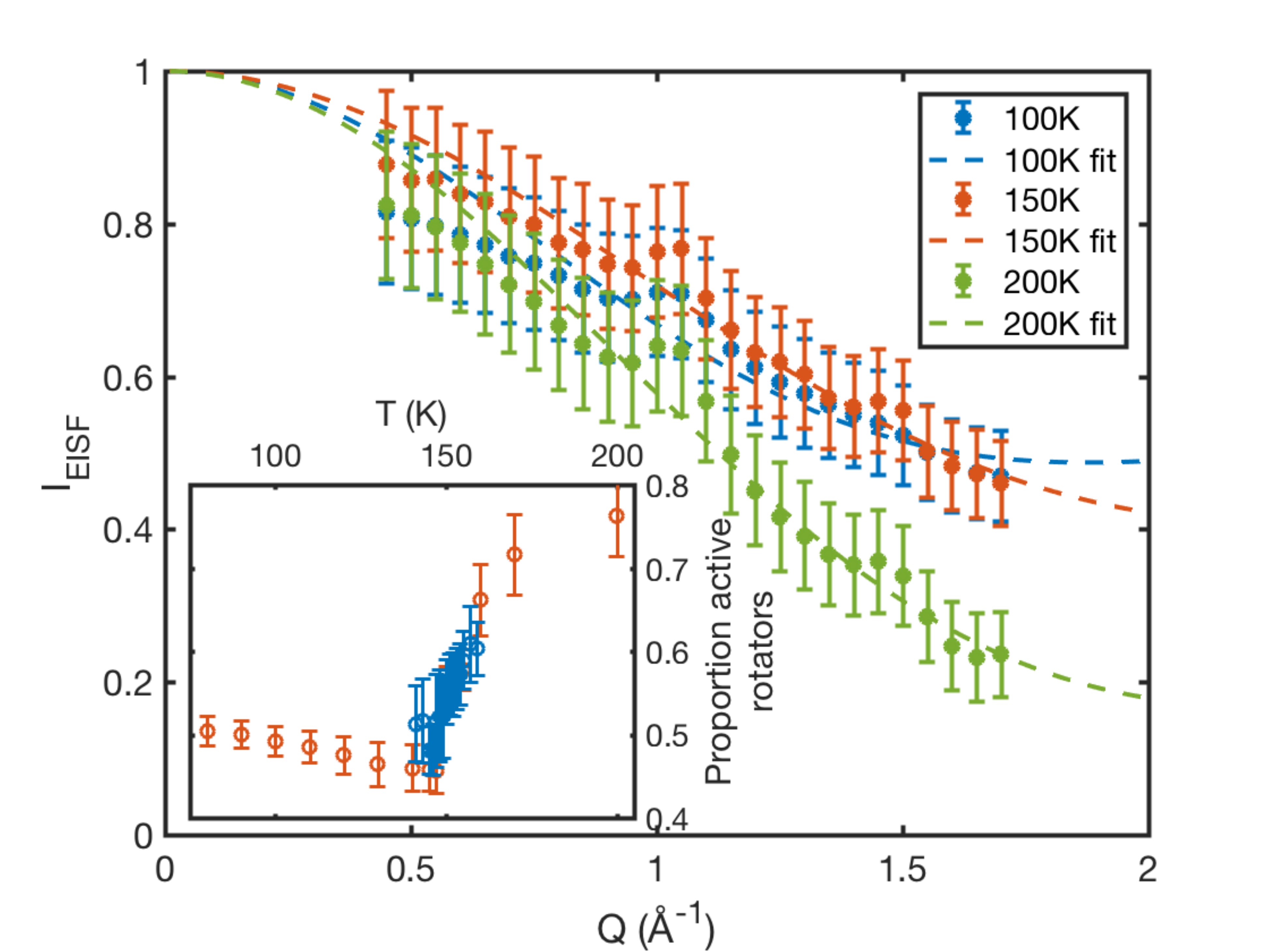}
\caption{\label{IEISF} The fits to the $I_{EISF}(Q)$ at 100 K, 150 K and 200 K, using the model of rotation on a spherical surface. The peak at $\sim$ 1 \AA$^{-1}$ is due to contamination from a PbBr$_{3}$ framework Bragg peak. Inset: the number of active rotators in the system with temperature. Blue and red data points are from two separate experiments. In the fit the radius of rotation was allowed to vary, and was found to be between 1.5 and 1.9 \AA, however, as this value is model dependent and sensitive to multiple phonon scattering, the temperature dependence of the radius of rotation is not plotted. A systematic error for multiple scattering has been included in I$_{EISF}$.}
\end{figure}

A plot of the $I_{EISF}(Q)$ extracted from the data at 150~K is shown in Fig. \ref{Lineshape2}$(b)$, with the fits from multiple models shown. The expected models to be appropriate based upon previous studies are the three, four and eight site jump models\cite{Leguy, Weller, Ren, Chen}. However, from the results obtained in our experiment, we conclude that the $Q$-range constrained by kinematics of the experiment is not sufficient to justify the choice of one model above another. Instead, the simplest model of a rotation on a spherical volume was used for our analysis, $f(Qr) = [j_{0}(Qr)]^{2}$, plotted in yellow in Fig. \ref{Lineshape2}$(b)$. 

A summary of the temperature dependence, based on this analysis, is plotted in Fig. \ref{IEISF}. In the main panel the fits for three temperatures are plotted showing that the model chosen is appropriate for all measurements across the temperature range. The inset to Fig. \ref{IEISF} shows how the proportion of rotators changes across the phases, and once again a clear increase can be seen at the 150~K phase transition. No increase is seen at the second phase transition, and the proportion of rotators begins to level out once more after 160 K. The radius of rotation was allowed to vary freely and was found to be between 1.5 and 1.9 \AA, suggesting that the motions seen here are of the C-N bond tumbling within the cage. From these results it is suggested that phase III has more in common dynamically with phase II than phase IV.

\subsection{High energy inelastic scattering data and Raman Studies}

Previous studies have already characterized the Raman spectra for members of the MAPbX$_{3}$ family, and the neutron spectra for MAPbI$_{3}$\cite{Leguy, Maaej, Pistor, Yaffe, Guo, Druzbiki}. While previous inelastic neutron scattering measurements on MAPbBr$_{3}$ focussed at low energies below $\sim$ 75 meV to characterize the soft modes~\cite{Swainson}, here we apply simultaneous neutron and Raman spectroscopy between 5 and 550 meV (40 to 4400~cm$^{-1}$) to investigate the internal molecular vibrations and their response to the structural transitions noted above.

\begin{figure}[t]
\includegraphics[width=8cm] {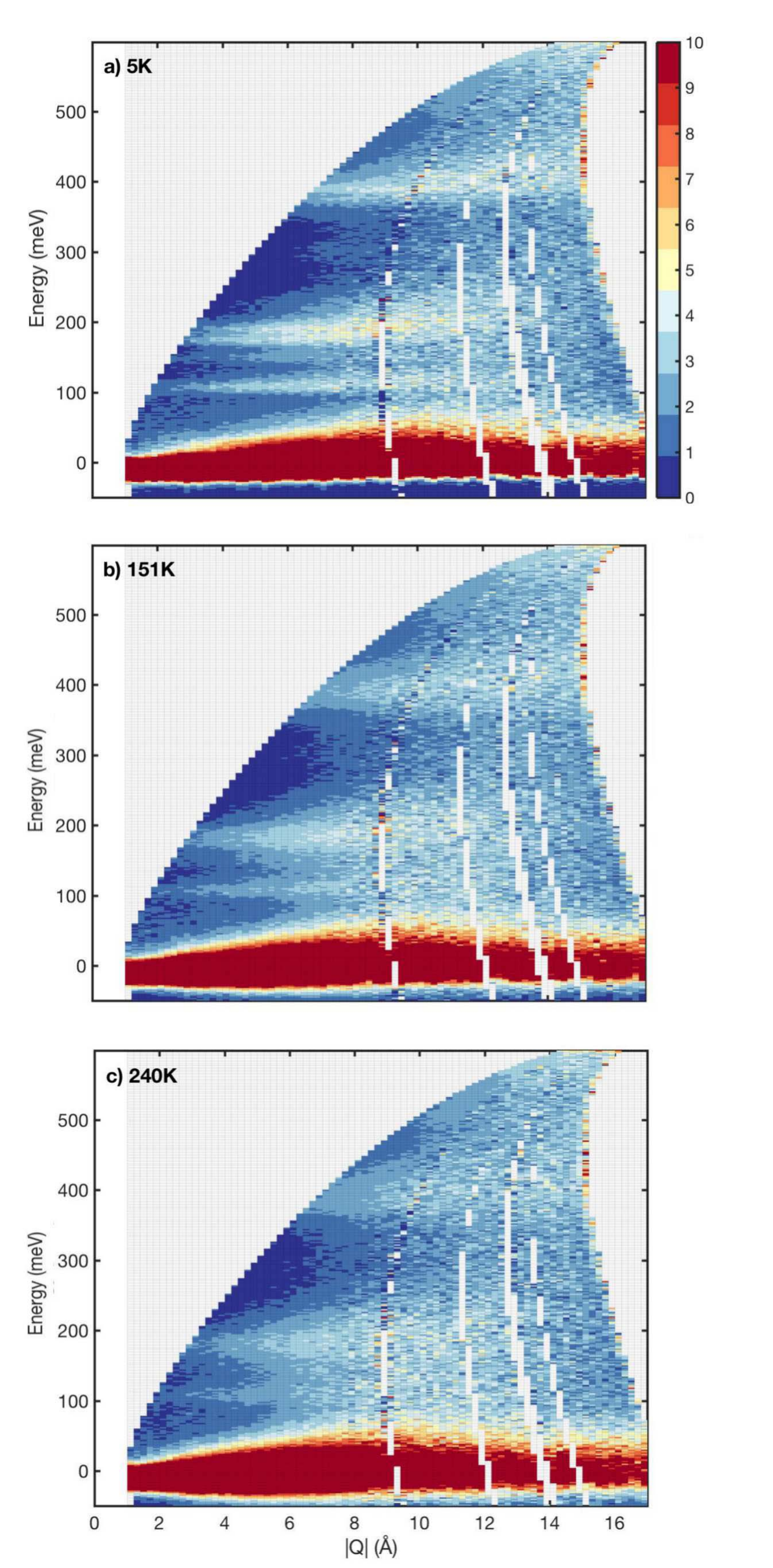}
\caption{\label{HydrogenRecoil} Slices taken from the 650 meV data at the three temperatures discussed in this paper, showing the $|Q|$ dependence of the intensities measured.}
\end{figure}

\begin{figure}[t]
\includegraphics[width=8cm] {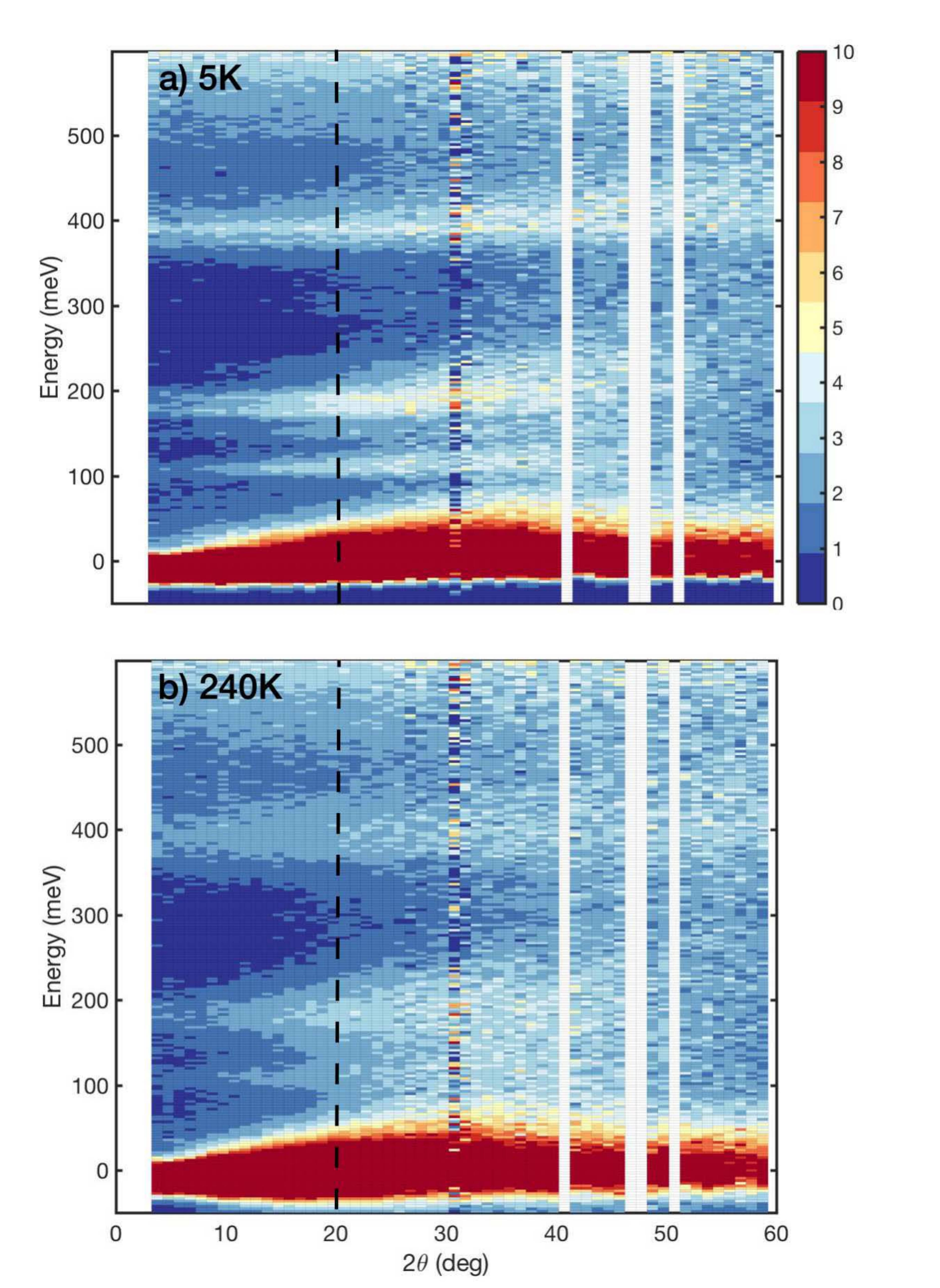}
\caption{\label{HydrogenRecoil2} Slices taken from the 650 meV data at base and high temperature, showing the $2\theta$ dependence of the intensities measured. The dashed black lines show the cutoff at 20$^{\circ}$ which was used to remove hydrogen recoil effects.}
\end{figure}

Typical momentum and energy data sets for E$_{i}$~=~650~meV obtained from MAPS are displayed in Fig. \ref{HydrogenRecoil} at 5, 151, and 240 K.  At small momentum transfers, well defined, underdamped in energy modes are observed.  This is particularly prominent in panel $(a)$ (5 K) where a series of sharp excitations are observed below $\sim$ 8 \AA$^{-1}$ and the intensity grows with increasing momentum transfer as expected for phonon modes or lattice excitations.  At higher momentum transfers, these excitations broaden in energy and even appear to disperse up in energy.  Given the large momentum transfers, in particular in comparison to Raman spectroscopy which is a strictly $|Q|$=0 probe, we speculate that this region is crossing over to the deep inelastic region where the impulse approximation applies and hydrogen recoil effects become important.

Hydrogen recoil is particularly prominent in neutron scattering due to the fact that the neutron mass is of the same order as that of the hydrogen nucleus, meaning that when the H nucleus is struck at high energies, it behaves as if it is free.  In the extreme limit of large energy and momentum transfers (as shown in Ref. \onlinecite{Stock} for polyethylene) the energy position of the hydrogen recoil scales as Q$^{2}$ with the maximum energy transfer occurring when $2\theta=90^{\circ}$.  In our data, hydrogen recoil effects appear as a broadening of the signal in the energy at larger momentum transfers, and scattering angle $2\theta$ and this is illustrated in Figs. \ref{HydrogenRecoil} (plotted as a function of $|Q|$) and  \ref{HydrogenRecoil2} (plotted as a function of $2\theta$) with an incident energy of E$_{i}$= 650 meV.   

In analyzing our neutron data and comparing it with Raman spectroscopy, a balance between minimizing contamination from recoil effects and obtaining enough statistics for a meaningful spectra had to be established.  We therefore chose to integrate the data in $2\theta$ in the range of 0$^{\circ}$ to 20$^{\circ}$ which was found not to result in broadening of the inelastic response, while providing enough statistics.  The region of integration is shown in Fig. \ref{HydrogenRecoil2}.  Having discussed how we obtained the neutron spectra, we now discuss the results and compare with Raman spectroscopy.

\begin{figure}[t]
\includegraphics[width=9cm] {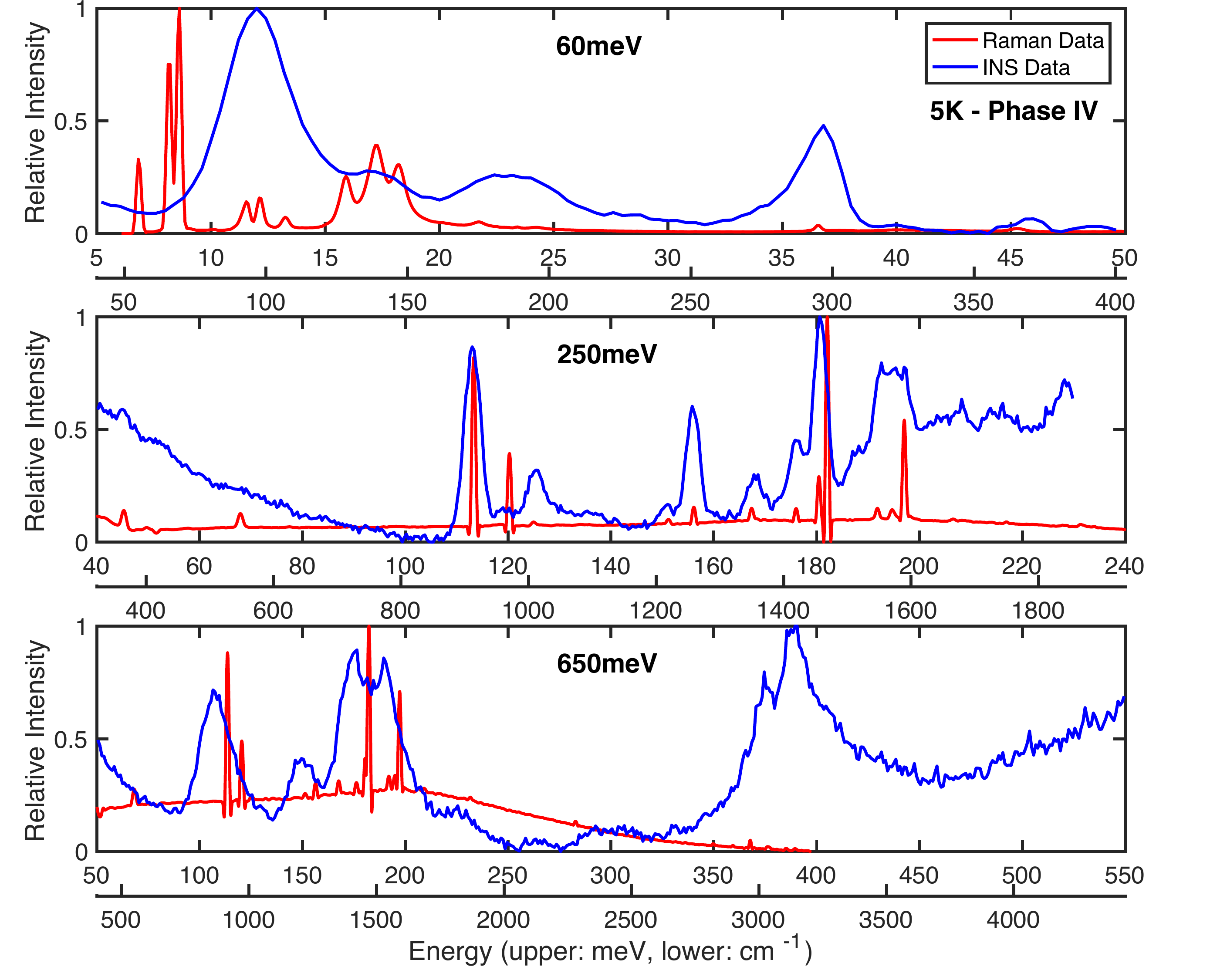}
\caption{\label{MAPSResults1} The results from MAPS at 60, 250 and 650 meV in the orthorhombic phase at base temperature (5.3 K). Raman data is shown in red; inelastic neutron scattering data in blue.}
\end{figure}

Figure \ref{MAPSResults1} show the data collected at base temperature (5 K in the orthorhombic phase) for both Raman and also neutron spectroscopy discussed above. The peak positions from the Raman data agree well with that found in previous studies, particularly that of Leguy \textit{et al.}, at base temperature, and as such we shall use their assignments of the Raman modes\cite{Leguy, Maaej, Pistor, Yaffe, Guo}. The peak positions in energy from Raman and neutron spectroscopy at high energies (E$_{i}$=250 and 650 meV) show good agreement, with the neutron spectroscopy data being broader in comparison as a result of instrumental resolution. However, there is a clear difference between the neutron spectroscopy and Raman spectra in the low energy area, a regime dominated by modes linked to the rotation of the PbBr$_{3}$ octahedra, and the lurching of the MA molecule~\cite{Niemann}. Such low-energy modes are highly dispersive throughout the Brillouin zone and, while Raman probes the modes at $|Q|$=0, the neutron spectra performs a momentum averaging $\tilde{S}(Q,E)={1\over {4\pi}} \int d\Omega S(\vec{Q},E)$ due to the powder nature of the sample.  With lattice vibrations that vary considerably with momentum, a larger difference is therefore expected between Raman and neutron spectroscopy.

The higher energy modes in Fig. \ref{MAPSResults1} are linked to intramolecular motions, such as: the stretching and breathing of intramolecular bonds; and the bending of the C-N bond.   Such motions are internal to the molecule and therefore do not disperse strongly with momentum.  All of the bending and stretching modes, as well as the mode for the breathing of the C-N bond, occur between 120~meV and 200 meV, whereas the C-H and N-H breathing modes are higher energy, occurring between 350 meV and 400 meV\cite{Leguy}.  These modes show good agreement between neutron and Raman techniques.

\begin{figure}[t]
\includegraphics[width=9cm] {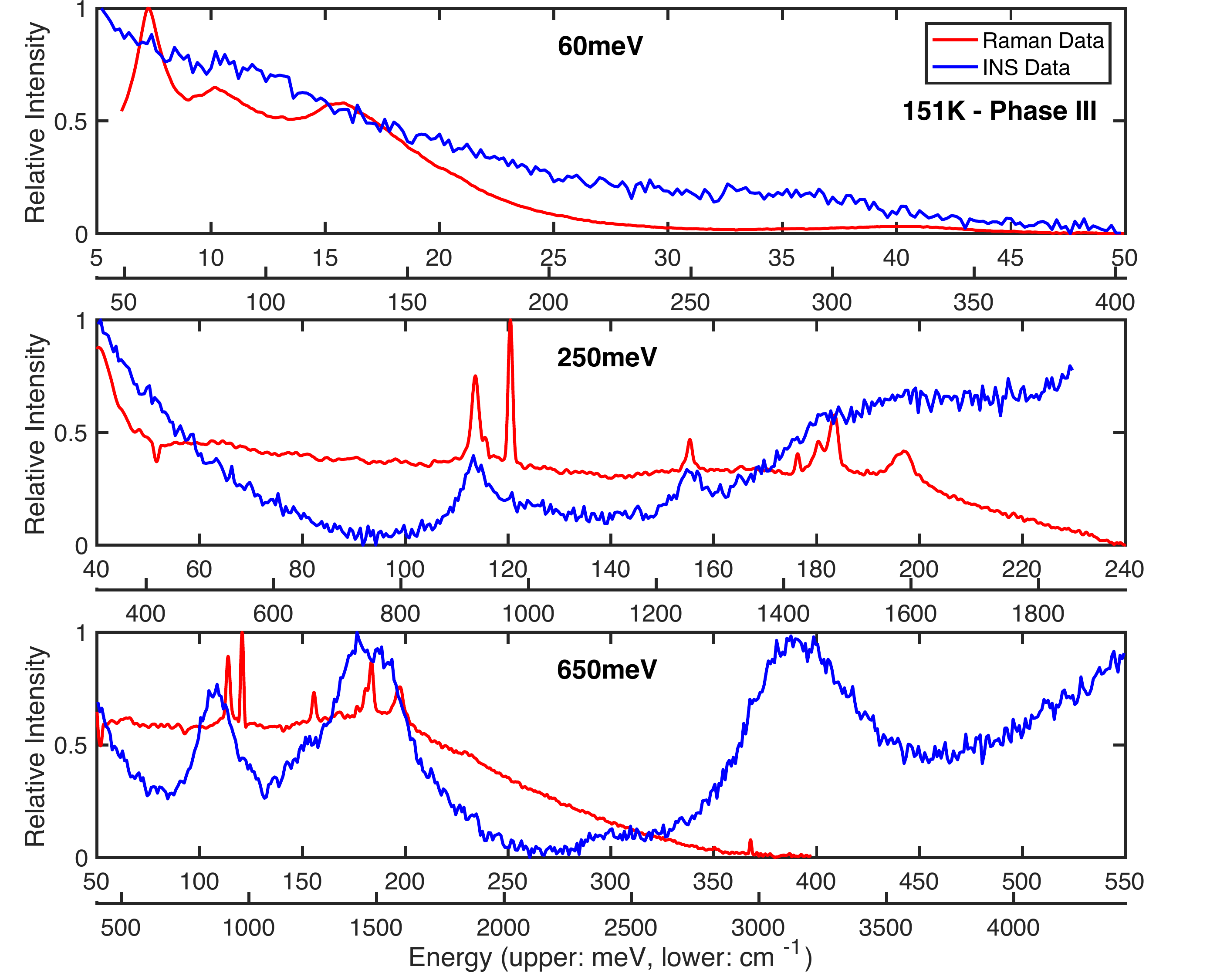}
\caption{\label{MAPSResults2} The results from MAPS at 60, 250 and 650 meV in the incommensurate/tetragonal phase (151 K).}
\end{figure}
\begin{figure}[t]
\includegraphics[width=9cm] {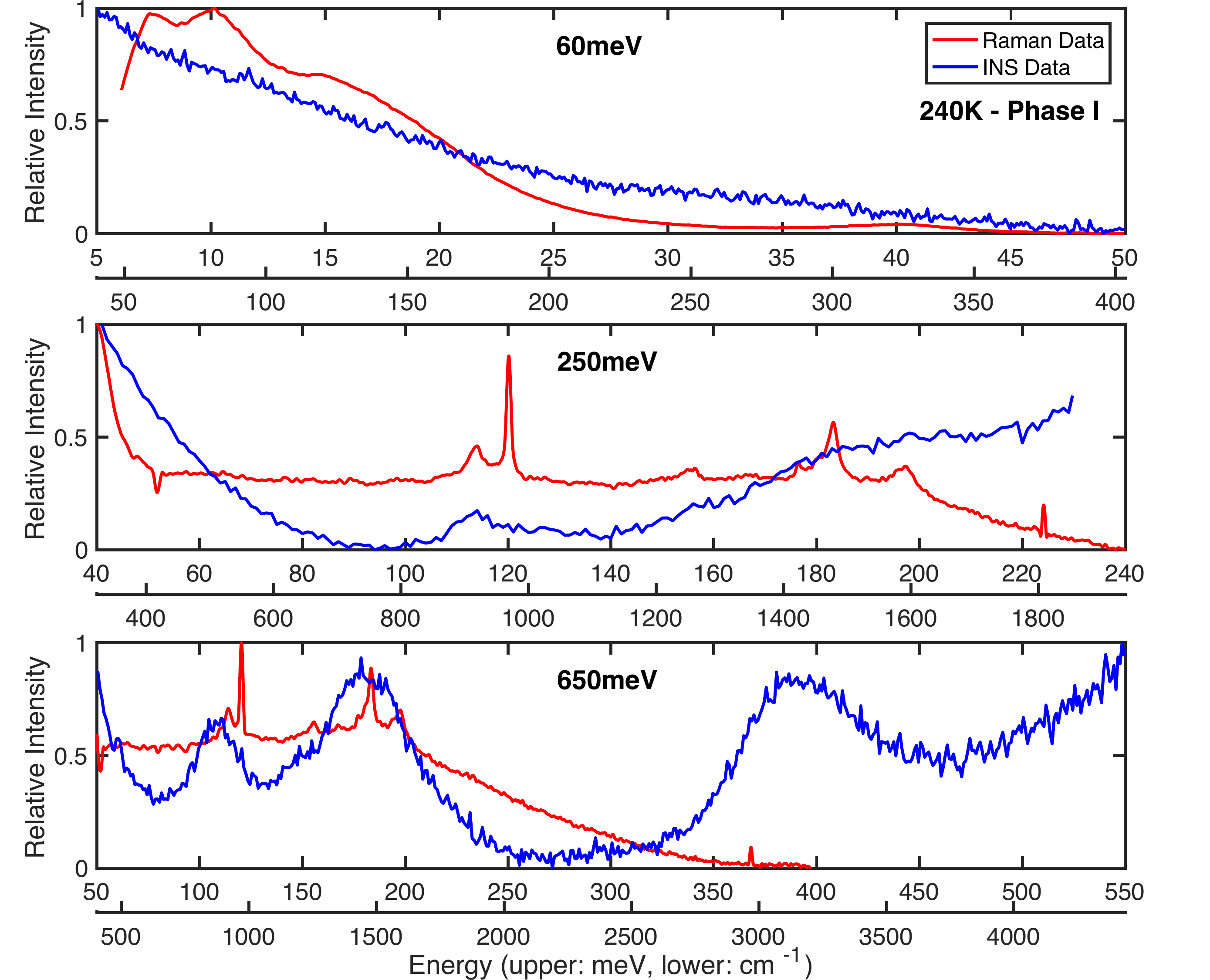}
\caption{\label{MAPSResults3} The results from MAPS at 60, 250 and 650 meV in the cubic phase (240 K).}
\end{figure}

Results at temperatures above the transition from an orthorhombic phase are shown in Figs. \ref{MAPSResults2} and \ref{MAPSResults3} for 151~K and 240 K respectively. When comparing this to the data taken at base temperature, a large broadening in energy of all excitations is observed in the neutron response indicative of a shortened lifetime.   Both Raman and neutron spectra show a large temporal broadening at low energies (below 50 meV) in agreement with previous neutron and Raman spectra~\cite{Druzbiki, Leguy}. This broadening was previously linked to the onset of fast molecular relaxational dynamics~\cite{Swainson}. These results, particularly those at energies below 60 meV, also confirm the faster relaxational timescale present at high temperatures postulated above based on high resolution quasielastic scattering, which lead to an increased flat background in the quasielastic scattering at temperatures greater than approximately 150~K.  We note that the fast molecular motions dominate the neutron cross section over modes associated with the inorganic cage~\cite{Niemann} at these temperatures owing to the dominant neutron cross section discussed above~\cite{Sears}.

For the higher energy modes, while the peak positions are in general agreement between neutron and Raman data, the temperature dependence of the linewidths is different.  The neutron and Raman results show a difference with the neutron response displaying a broadening with increased temperature and also a broadening over Raman spectra taken over a comparable energy range.    This is illustrated in Table \ref{FWHMtable} which displays the temperature variation of the full width  at half maximum for the peak at 113 meV (915 cm$^{-1}$).  This peak was chosen due to it's clear separation from other peaks at  all temperatures, and has been identified to be linked to the rocking motions of the MA cation~\cite{Niemann}. Both neutron and Raman peaks at low temperature are resolution limited.  The peak measured with neutrons broadens to $\sim$ 10 meV while the Raman peak broadens only to $\sim$ 3 meV.  

\begin{figure}[t]
\includegraphics[width=8cm] {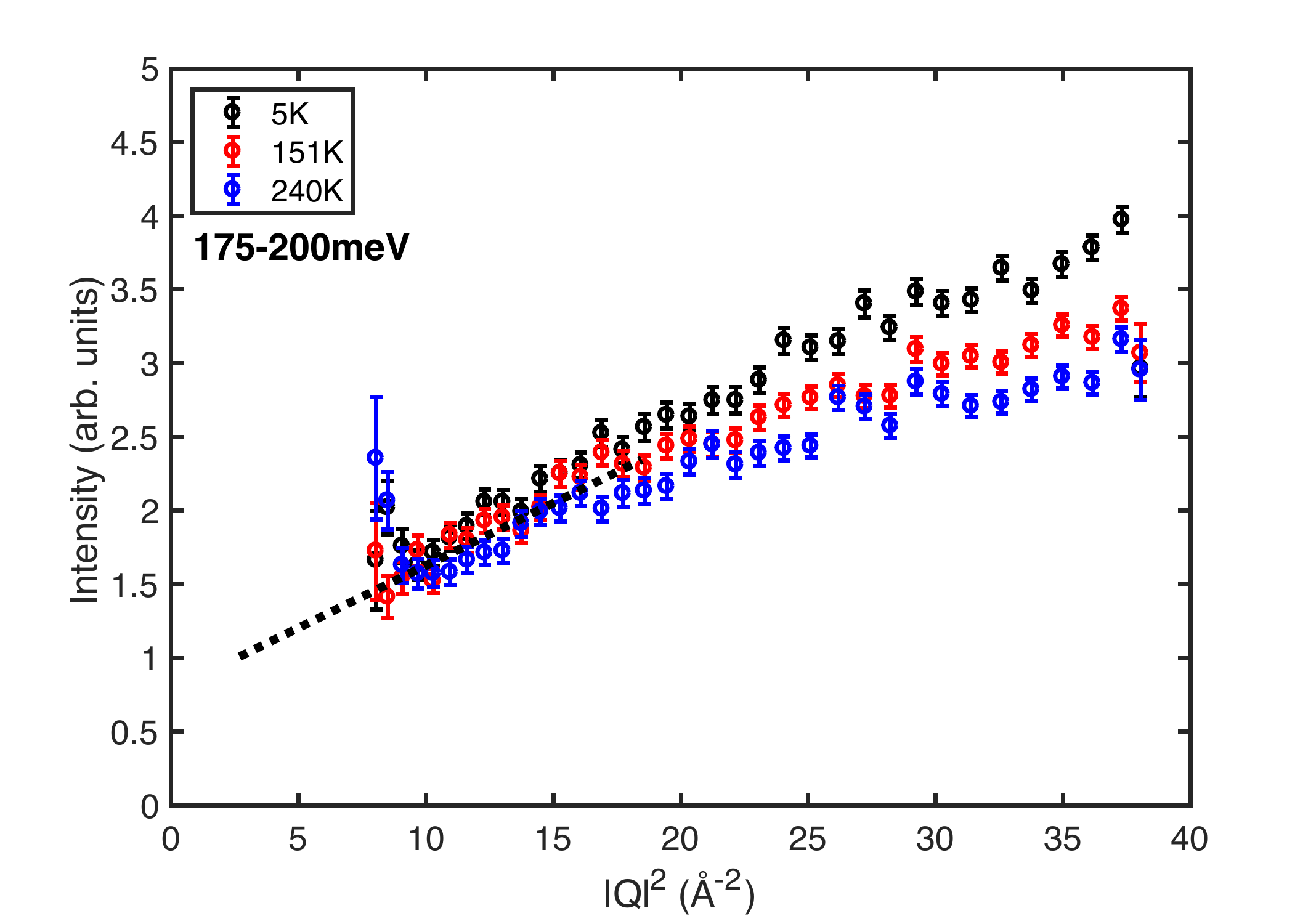}
\caption{\label{MultiplePhonon} Data collected at 650 meV, integrated over an energy range of 175-200meV and plotted against $|Q|^{2}$. A dotted line is plotted as a guide to the eye.}
\end{figure}

One reason for this increased broadening in the neutron response could be due to the possibility of multiple phonon scattering which effectively folds in scattering from larger scattering angles resulting from the large neutron cross section of hydrogen.   We investigate the possibility of such an effect in Fig. \ref{MultiplePhonon} where we plot the momentum dependence of the neutron response over the range of 175-200 meV as a function of $Q^{2}$.  This energy range has been chosen for presentation purposes due to the presence of a well defined peak in the 650 meV data at all temperatures probed.  An increase in multiple phonon scattering would manifest as differing $y$-intercepts at each temperature.  The data is in good agreement at low momentum transfers indicating no observable enhancement or change of multiple phonon scattering with increased temperature.   Normally one would expect the $y$-intercept in all cases to be zero, however here no explicit background subtraction has been carried out, leading to the non-zero value extrapolated here.  The plot in Fig. \ref{MultiplePhonon} is selected for a particular and representative energy range.  However, all other peaks were observed to show a lack of a temperature dependence in the $\lim_{Q\rightarrow 0}$ as shown here.

\begin{table}
\caption{The full width at half maximum of the peak at 113meV  
(915cm$^{-1}$) for three temperatures.}
\label{FWHMtable}
\begin{tabular}{ |c||c|c|c|c| }
\hline
\multicolumn{1}{|c||}{} & \multicolumn{3}{c|}{FWHM of 113meV peak (meV)} \\
\hline
Temperature & 5.3 K & 151 K & 240 K \\
\hline
Neutron & $5.122 \pm 0.151$ &  $8.253 \pm 0.968$ &  $10.176 \pm 2.872$ \\
Raman &  $1.057 \pm 0.019$ &  $2.474 \pm 0.040$ &  $2.710 \pm 0.090$ \\
\hline
\end{tabular}
\end{table}

There are several possible reasons why this difference between neutron and Raman techniques might be observed. Firstly, Raman spectra are strictly at $|Q| = 0$, whereas inelastic neutron scattering integrates and averages over all $|Q|$ as discussed above owing to kinematics and powder averaging.  This means that an effect at other positions in momentum, for example the zone boundary,  could cause the momentum averaged energy width to be increased more than what is measured at a single wavevector like with Raman spectroscopy at $\vec{Q}$=0. Secondly, the inelastic neutron spectra are dominated by hydrogen, whereas Raman is more sensitive to heavier elements, meaning that neutron scattering disproportionately samples the motions on the hydrogen sites. This is further confirmed by the sensitivity of these high energy excitations to recoil effects at large momentum transfers and scattering angles as discussed above.  We discuss these two possibilities in more detail below in the discussion section.

There is the possibility Raman techniques preferentially measure molecules close to the surface (see for example surface enhanced spectroscopy discussed in Ref. \onlinecite{Chen2}), whereas inelastic neutron scattering considers the bulk system. However, while surface over bulk differences have been suggested in the organic-inorganic perovskites (for example the effect of dimensionality discussed in Ref. \onlinecite{Quan}), this latter possibility is unlikely in this case given that surface-enhanced Raman spectroscopy in molecular systems and liquids has typically resulted in a broadening of linewidths or intensity enhancement at the surface.  We observe the reverse case with the Raman measurements being more well defined in energy than the bulk neutron scattering technique.  The agreement between neutron and Raman at lower energies also does not suggest a near surface versus bulk difference for neutron and Raman spectra noted here.

\section{Discussion}

We have presented results from both a neutron low-energy and high resolution quasielastic scattering study and high energy neutron spectroscopy measurements.  The low energy neutron experiments performed on IRIS probes the single atom motions with a range of energy scales whereas the higher energy inelastic data, obtained on MAPS and compared with Raman spectroscopy, measures harmonic modes of the molecules.  We have focussed on the internal molecular motions at energies above $\sim$ 100 meV.  

The quasielastic neutron scattering from IRIS is unable to provide any conclusive information on the exact molecular jump model given the limited range in momentum forced by kinematics of neutron scattering.  However, by applying arguably the simplest model of free motion on a sphere we observed changes in the quasielastic linewidth at the 150 K phase transition, and the appearance of the second, broader in energy contribution at higher temperatures.  This indicates  a rotational mode active in the low temperature phase and a second faster energy scale that becomes active at the phase transition to the tetragonal phase\cite{Leguy, Chen}.   We also observe a precipitous increase in the number of active molecules at $\sim$ 150 K indicative of an increase in local molecular disorder as temperature is increased.  This is corroborated by diffraction studies\cite{Ren, Whitfield, Mashiyama}, NMR studies\cite{Roiland} and computational investigations\cite{Mattoni, Lee}. 

A similar result is obtained from our spectroscopic study of the internal molecular motions obtained using a combination of the inelastic neutron scattering and Raman spectroscopy on MAPS.  The energy locations and the temperature dependence of the Raman data presented are generally in agreement with previous studies\cite{Leguy, Druzbiki, Maaej, Pistor, Yaffe, Guo}.  At low energies below $\sim$ 50 meV, Raman and neutron spectroscopy are in agreement with both showing a significant temporal broadening of the excitations upon heating from the orthorhombic phase, indicative of faster fluctuations with shortened lifetimes. This is corroborated by the quasielastic neutron data probing molecular motions discussed above. However, Raman and neutron spectroscopy show differences at higher energies with neutron spectroscopy measuring broadened linewidths of internal molecular motions.  This broadening is indicative of an increased distribution of energy scales for these motions.

As discussed above, this difference can be explained by either noting that neutron spectroscopy is preferentially sensitive to hydrogen motions given the large cross section over other heavier atoms or the averaging over all momentum in contrast to the fact that Raman only probes $|Q|$=0.  Given that electronic bandstructure may indicate a preferential coupling between lattice and electronic degrees of freedom at non zero $|Q|$~\cite{Brivio,Motta}, it maybe expected that there might be a coupling of these internal modes at regions away from the zone center.  However, these high energy internal molecular excitations are associated with excitations which are highly localized in real space and therefore are expected to show no momentum dependence.  We therefore associate the difference between Raman and neutron results at high temperatures due to the dominance of the hydrogen cross section in neutron spectroscopy.  This would indicate a distribution of crystalline electric fields for the hydrogen resulting in a corresponding spread in energy scales for the molecular motions involving them.  Theoretically, such a scenario has been predicted as outlined in Ref. \onlinecite{Even}, and the importance of hydrogen bonding within this system has been previously discussed in Ref. \onlinecite{Lee}, in line with our findings. 

The suggestion that the molecular cation in the system is experiencing a range of different local environments is supported in other experimental studies: ARPES studies of MAPbBr$_{3}$ find both centrosymmetric and non-centrosymmetric domains in the crystal, which persist throughout the low temperature phase\cite{Niesner}. This effect has also been reported for other phases of MAPbBr$_{3}$ and MAPbI$_{3}$, which reports that symmetry breaking domains survive past the phase transition between the cubic and tetragonal phases both when heated and cooled\cite{Comin, Even2}. Additionally, significant disorder on the halide site has also been reported in diffraction results for MAPbI$_{3}$~\cite{Minns}; such disorder may influence local hydrogen crystalline electric field environments.

It is clear that the onset of molecular dynamics at 150~K results in a large scale dynamic molecular disorder of the entire crystal. This is evidenced from both the onset of dynamic sites found with low-energy quasielastic neutron scattering and also the presence of a distribution of environments for the hydrogen site.  This has been  previously discussed theoretically, with the system being compared to plastic crystals\cite{Even}, and the term ``glassy disorder" was coined to describe the molecular motions in the low temperature phase\cite{Fabini}. However, the present results imply that the change in dynamics between the orthorhombic and tetragonal phases is more akin to a orientational melting analogous to that in the smectic to nematic transition in liquid crystals and, more precisely, the order-disorder transitions reported in minerals such as calcite\cite{Bell}. 

We shall now discuss the implications of these findings on the electronic properties of this material. The key contributions to the edges of the band structure come from the hybridization of the $5p$ orbitals of iodine and $6s$ orbitals of lead for the valence band, and the empty $6p$ orbitals of lead for the conduction band\cite{Endres, Herz}. While the MA cation does not contribute directly to the band structure, the coupling between the motion of the molecule and the structure of the inorganic components makes it a key influence upon the electronic response, and hence upon the photovoltaic properties. 

Therefore, the observed difference in the dynamics of the system between low and high temperature structural phases implies that there will be a linked change in the electronic and photovoltaic properties. Photoluminescence spectra results show that there is a change in the energy of the maximum photoluminescence intensity at the 150 K transition\cite{Wright}, in agreement with the observation made here that the key change in whole body molecular dynamics is at this temperature. This suggests that it is the whole body rotations of the MA cation and their interactions with the surrounding inorganic framework~\cite{Gao} that is key to understanding the photovoltaic properties. However, it is important to note that the change in photoluminescence observed at 150~K is not large.

Further to these observations of the photovoltaic properties, studies have also identified a coupling between the cation rotation to the quasiparticle band gap of the system, through hydrogen bonding between the cation and inorganic cages. The hydrogen bonding has been identified as strongly linked with the octahedral tilts of the system, and the phase transitions within the system, with the hydrogen bonding stronger in the low-T phase where the molecular motion is lower\cite{Lee, Lee2, Ong, Butler}. Due to the fact that the structure is so heavily dependent on the cation dynamics, this then influences the band gap of the system, with the states near the top of the valence band stabilized by the octahedral tilting in the low temperature phase with fewer cation dynamics\cite{Gao, Lee, Lee2}. 

Additional evidence showing that the coupling between molecular motions and electronic properties is relevant for the photovoltaic properties, it has also been found that the cation dynamics provides screening which protects the energetic carriers\cite{Zhu}. This implies that, from the results presented here, the assumption can be made that the low temperature phase has least carrier protection and thus the carriers are less long-lived.  Also, a large Rashba splitting~\cite{Niesner} resulting from a breaking of inversion symmetry and strong spin orbit coupling~\cite{Even2} has been reported. Our results suggest that it is not only the lead halide framework coupled to electronic properties~\cite{Wright}, but also the molecular motions.

\section{Conclusion}

In conclusion, we have reported two neutron scattering studies which show that there is a significant change in the molecular dynamics of the MA cation in MAPbBr$_{3}$ as temperature is increased from the orthorhombic phase to the higher temperature structural phases.  We have used quasielastic neutron scattering data to establish that whole body molecular dynamics are activated at 150~K.  From a comparison of Raman and high energy inelastic neutron scattering and Raman, the significance of the local environment around the hydrogen sites is implicated pointing to the importance of hydrogen bonding to the properties of the system. 
It is known that the motion of the MA cation is able to strongly influence the photovoltaic properties through interactions with the surrounding PbBr$_{3}$ octahedra\cite{Herz, Gao, Zhu}. The onset of rotational motion agrees with an observed feature in the photoluminescence data\cite{Wright}, supporting the observation that the behavior of the organic cation is key to understanding the photovoltaic and optic properties of this compound.

\section*{Acknowledgements}

Support is gratefully acknowledged from the EPSRC, the STFC, the Royal Society, and the Carnegie Trust for the Universities of Scotland. We thank P. M. Gehring, I. P. Swainson, J. P. Attfield, and M. Songvilay for helpful discussions.

\bibliographystyle{aipnum4-1}

\begin{thebibliography}{62}%
\makeatletter
\providecommand \@ifxundefined [1]{%
 \@ifx{#1\undefined}
}%
\providecommand \@ifnum [1]{%
 \ifnum #1\expandafter \@firstoftwo
 \else \expandafter \@secondoftwo
 \fi
}%
\providecommand \@ifx [1]{%
 \ifx #1\expandafter \@firstoftwo
 \else \expandafter \@secondoftwo
 \fi
}%
\providecommand \natexlab [1]{#1}%
\providecommand \enquote  [1]{``#1''}%
\providecommand \bibnamefont  [1]{#1}%
\providecommand \bibfnamefont [1]{#1}%
\providecommand \citenamefont [1]{#1}%
\providecommand \href@noop [0]{\@secondoftwo}%
\providecommand \href [0]{\begingroup \@sanitize@url \@href}%
\providecommand \@href[1]{\@@startlink{#1}\@@href}%
\providecommand \@@href[1]{\endgroup#1\@@endlink}%
\providecommand \@sanitize@url [0]{\catcode `\\12\catcode `\$12\catcode
  `\&12\catcode `\#12\catcode `\^12\catcode `\_12\catcode `\%12\relax}%
\providecommand \@@startlink[1]{}%
\providecommand \@@endlink[0]{}%
\providecommand \url  [0]{\begingroup\@sanitize@url \@url }%
\providecommand \@url [1]{\endgroup\@href {#1}{\urlprefix }}%
\providecommand \urlprefix  [0]{URL }%
\providecommand \Eprint [0]{\href }%
\providecommand \doibase [0]{http://dx.doi.org/}%
\providecommand \selectlanguage [0]{\@gobble}%
\providecommand \bibinfo  [0]{\@secondoftwo}%
\providecommand \bibfield  [0]{\@secondoftwo}%
\providecommand \translation [1]{[#1]}%
\providecommand \BibitemOpen [0]{}%
\providecommand \bibitemStop [0]{}%
\providecommand \bibitemNoStop [0]{.\EOS\space}%
\providecommand \EOS [0]{\spacefactor3000\relax}%
\providecommand \BibitemShut  [1]{\csname bibitem#1\endcsname}%
\let\auto@bib@innerbib\@empty
\bibitem [{\citenamefont {Saparov}\ and\ \citenamefont
  {Mitzi}(2016)}]{Saparov}%
  \BibitemOpen
  \bibfield  {author} {\bibinfo {author} {\bibfnamefont {B.}~\bibnamefont
  {Saparov}}\ and\ \bibinfo {author} {\bibfnamefont {D.~B.}\ \bibnamefont
  {Mitzi}},\ }\href {\doibase 10.1021/acs.chemrev.5b00715} {\bibfield
  {journal} {\bibinfo  {journal} {Chem. Rev.}\ }\textbf {\bibinfo {volume}
  {116}},\ \bibinfo {pages} {4558} (\bibinfo {year} {2016})}\BibitemShut
  {NoStop}%
\bibitem [{\citenamefont {Park}\ \emph {et~al.}(2015)\citenamefont {Park},
  \citenamefont {Choi}, \citenamefont {Yan}, \citenamefont {Yang},
  \citenamefont {Luther}, \citenamefont {Wei}, \citenamefont {Parilla},\ and\
  \citenamefont {Zhu}}]{Park}%
  \BibitemOpen
  \bibfield  {author} {\bibinfo {author} {\bibfnamefont {J.-S.}\ \bibnamefont
  {Park}}, \bibinfo {author} {\bibfnamefont {S.}~\bibnamefont {Choi}}, \bibinfo
  {author} {\bibfnamefont {Y.}~\bibnamefont {Yan}}, \bibinfo {author}
  {\bibfnamefont {Y.}~\bibnamefont {Yang}}, \bibinfo {author} {\bibfnamefont
  {J.~M.}\ \bibnamefont {Luther}}, \bibinfo {author} {\bibfnamefont {S.-H.}\
  \bibnamefont {Wei}}, \bibinfo {author} {\bibfnamefont {P.}~\bibnamefont
  {Parilla}}, \ and\ \bibinfo {author} {\bibfnamefont {K.}~\bibnamefont
  {Zhu}},\ }\href {\doibase 10.1021/acs.jpclett.5b01699} {\bibfield  {journal}
  {\bibinfo  {journal} {J. Phys. Chem. Lett.}\ }\textbf {\bibinfo {volume}
  {6}},\ \bibinfo {pages} {4304} (\bibinfo {year} {2015})}\BibitemShut
  {NoStop}%
\bibitem [{\citenamefont {Noh}\ \emph {et~al.}(2013)\citenamefont {Noh},
  \citenamefont {Im}, \citenamefont {Heo}, \citenamefont {Mandal},\ and\
  \citenamefont {Seok}}]{Noh}%
  \BibitemOpen
  \bibfield  {author} {\bibinfo {author} {\bibfnamefont {J.~H.}\ \bibnamefont
  {Noh}}, \bibinfo {author} {\bibfnamefont {S.~H.}\ \bibnamefont {Im}},
  \bibinfo {author} {\bibfnamefont {J.~H.}\ \bibnamefont {Heo}}, \bibinfo
  {author} {\bibfnamefont {T.~N.}\ \bibnamefont {Mandal}}, \ and\ \bibinfo
  {author} {\bibfnamefont {S.~I.}\ \bibnamefont {Seok}},\ }\href {\doibase
  10.1021/nl400349b} {\bibfield  {journal} {\bibinfo  {journal} {Nano Lett.}\
  }\textbf {\bibinfo {volume} {13}},\ \bibinfo {pages} {1764} (\bibinfo {year}
  {2013})}\BibitemShut {NoStop}%
\bibitem [{\citenamefont {Zhou}\ \emph {et~al.}(2014)\citenamefont {Zhou},
  \citenamefont {Chen}, \citenamefont {Li}, \citenamefont {Luo}, \citenamefont
  {Song}, \citenamefont {Duan}, \citenamefont {Hong}, \citenamefont {You},
  \citenamefont {Liu},\ and\ \citenamefont {Yang}}]{Zhou}%
  \BibitemOpen
  \bibfield  {author} {\bibinfo {author} {\bibfnamefont {H.}~\bibnamefont
  {Zhou}}, \bibinfo {author} {\bibfnamefont {Q.}~\bibnamefont {Chen}}, \bibinfo
  {author} {\bibfnamefont {G.}~\bibnamefont {Li}}, \bibinfo {author}
  {\bibfnamefont {S.}~\bibnamefont {Luo}}, \bibinfo {author} {\bibfnamefont
  {T.-b.}\ \bibnamefont {Song}}, \bibinfo {author} {\bibfnamefont {H.-S.}\
  \bibnamefont {Duan}}, \bibinfo {author} {\bibfnamefont {Z.}~\bibnamefont
  {Hong}}, \bibinfo {author} {\bibfnamefont {J.}~\bibnamefont {You}}, \bibinfo
  {author} {\bibfnamefont {Y.}~\bibnamefont {Liu}}, \ and\ \bibinfo {author}
  {\bibfnamefont {Y.}~\bibnamefont {Yang}},\ }\href {\doibase
  10.1126/science.1254050} {\bibfield  {journal} {\bibinfo  {journal}
  {Science}\ }\textbf {\bibinfo {volume} {345}},\ \bibinfo {pages} {542}
  (\bibinfo {year} {2014})}\BibitemShut {NoStop}%
\bibitem [{\citenamefont {Herz}(2016)}]{Herz}%
  \BibitemOpen
  \bibfield  {author} {\bibinfo {author} {\bibfnamefont {L.~M.}\ \bibnamefont
  {Herz}},\ }\href {\doibase 10.1146/annurev-physchem-040215-112222} {\bibfield
   {journal} {\bibinfo  {journal} {Annual Review of Physical Chemistry}\
  }\textbf {\bibinfo {volume} {67}},\ \bibinfo {pages} {65} (\bibinfo {year}
  {2016})}\BibitemShut {NoStop}%
\bibitem [{\citenamefont {Bennett}\ \emph {et~al.}(2016)\citenamefont
  {Bennett}, \citenamefont {Cheetham}, \citenamefont {Fuchs},\ and\
  \citenamefont {Coudert}}]{Bennett}%
  \BibitemOpen
  \bibfield  {author} {\bibinfo {author} {\bibfnamefont {T.~D.}\ \bibnamefont
  {Bennett}}, \bibinfo {author} {\bibfnamefont {A.~K.}\ \bibnamefont
  {Cheetham}}, \bibinfo {author} {\bibfnamefont {A.~H.}\ \bibnamefont {Fuchs}},
  \ and\ \bibinfo {author} {\bibfnamefont {F.-X.}\ \bibnamefont {Coudert}},\
  }\href {http://dx.doi.org/10.1038/nchem.2691} {\bibfield  {journal} {\bibinfo
   {journal} {Nat. Chem.}\ }\textbf {\bibinfo {volume} {9}},\ \bibinfo {pages}
  {11 EP } (\bibinfo {year} {2016})}\BibitemShut {NoStop}%
\bibitem [{\citenamefont {Poglitsch}\ and\ \citenamefont
  {Weber}(1987)}]{Poglitsch}%
  \BibitemOpen
  \bibfield  {author} {\bibinfo {author} {\bibfnamefont {A.}~\bibnamefont
  {Poglitsch}}\ and\ \bibinfo {author} {\bibfnamefont {D.}~\bibnamefont
  {Weber}},\ }\href {\doibase 10.1063/1.453467} {\bibfield  {journal} {\bibinfo
   {journal} {J. Chem. Phys.}\ }\textbf {\bibinfo {volume} {87}},\ \bibinfo
  {pages} {6373} (\bibinfo {year} {1987})}\BibitemShut {NoStop}%
\bibitem [{\citenamefont {Brock}\ \emph
  {et~al.}(1989{\natexlab{a}})\citenamefont {Brock}, \citenamefont {Noh},
  \citenamefont {McClain}, \citenamefont {Litster}, \citenamefont {Birgeneau},
  \citenamefont {Aharony}, \citenamefont {Horn},\ and\ \citenamefont
  {Liang}}]{Brock}%
  \BibitemOpen
  \bibfield  {author} {\bibinfo {author} {\bibfnamefont {J.~D.}\ \bibnamefont
  {Brock}}, \bibinfo {author} {\bibfnamefont {D.~Y.}\ \bibnamefont {Noh}},
  \bibinfo {author} {\bibfnamefont {B.~R.}\ \bibnamefont {McClain}}, \bibinfo
  {author} {\bibfnamefont {J.~D.}\ \bibnamefont {Litster}}, \bibinfo {author}
  {\bibfnamefont {R.~J.}\ \bibnamefont {Birgeneau}}, \bibinfo {author}
  {\bibfnamefont {A.}~\bibnamefont {Aharony}}, \bibinfo {author} {\bibfnamefont
  {P.~M.}\ \bibnamefont {Horn}}, \ and\ \bibinfo {author} {\bibfnamefont
  {J.~C.}\ \bibnamefont {Liang}},\ }\href {\doibase 10.1007/BF01307386}
  {\bibfield  {journal} {\bibinfo  {journal} {Z. Phys. B Cond. Matt.}\ }\textbf
  {\bibinfo {volume} {74}},\ \bibinfo {pages} {197} (\bibinfo {year}
  {1989}{\natexlab{a}})}\BibitemShut {NoStop}%
\bibitem [{\citenamefont {Ren}\ \emph {et~al.}(2016)\citenamefont {Ren},
  \citenamefont {Oswald}, \citenamefont {Wang}, \citenamefont {McCandless},\
  and\ \citenamefont {Chan}}]{Ren}%
  \BibitemOpen
  \bibfield  {author} {\bibinfo {author} {\bibfnamefont {Y.}~\bibnamefont
  {Ren}}, \bibinfo {author} {\bibfnamefont {I.~W.~H.}\ \bibnamefont {Oswald}},
  \bibinfo {author} {\bibfnamefont {X.}~\bibnamefont {Wang}}, \bibinfo {author}
  {\bibfnamefont {G.~T.}\ \bibnamefont {McCandless}}, \ and\ \bibinfo {author}
  {\bibfnamefont {J.~Y.}\ \bibnamefont {Chan}},\ }\href {\doibase
  10.1021/acs.cgd.6b00297} {\bibfield  {journal} {\bibinfo  {journal} {Cryst.
  Growth Des.}\ }\textbf {\bibinfo {volume} {16}},\ \bibinfo {pages} {2945}
  (\bibinfo {year} {2016})}\BibitemShut {NoStop}%
\bibitem [{\citenamefont {Mashiyama}\ \emph {et~al.}(2007)\citenamefont
  {Mashiyama}, \citenamefont {Kawamura}, \citenamefont {Kasano}, \citenamefont
  {Asahi}, \citenamefont {Noda},\ and\ \citenamefont {Kimura}}]{Mashiyama}%
  \BibitemOpen
  \bibfield  {author} {\bibinfo {author} {\bibfnamefont {H.}~\bibnamefont
  {Mashiyama}}, \bibinfo {author} {\bibfnamefont {Y.}~\bibnamefont {Kawamura}},
  \bibinfo {author} {\bibfnamefont {H.}~\bibnamefont {Kasano}}, \bibinfo
  {author} {\bibfnamefont {T.}~\bibnamefont {Asahi}}, \bibinfo {author}
  {\bibfnamefont {Y.}~\bibnamefont {Noda}}, \ and\ \bibinfo {author}
  {\bibfnamefont {H.}~\bibnamefont {Kimura}},\ }\href {\doibase
  10.1080/00150190701196435} {\bibfield  {journal} {\bibinfo  {journal}
  {Ferroelectrics}\ }\textbf {\bibinfo {volume} {348}},\ \bibinfo {pages} {182}
  (\bibinfo {year} {2007})}\BibitemShut {NoStop}%
\bibitem [{\citenamefont {Whitfield}\ \emph {et~al.}(2016)\citenamefont
  {Whitfield}, \citenamefont {Herron}, \citenamefont {Guise}, \citenamefont
  {Page}, \citenamefont {Cheng}, \citenamefont {Milas},\ and\ \citenamefont
  {Crawford}}]{Whitfield}%
  \BibitemOpen
  \bibfield  {author} {\bibinfo {author} {\bibfnamefont {P.~S.}\ \bibnamefont
  {Whitfield}}, \bibinfo {author} {\bibfnamefont {N.}~\bibnamefont {Herron}},
  \bibinfo {author} {\bibfnamefont {W.~E.}\ \bibnamefont {Guise}}, \bibinfo
  {author} {\bibfnamefont {K.}~\bibnamefont {Page}}, \bibinfo {author}
  {\bibfnamefont {Y.~Q.}\ \bibnamefont {Cheng}}, \bibinfo {author}
  {\bibfnamefont {I.}~\bibnamefont {Milas}}, \ and\ \bibinfo {author}
  {\bibfnamefont {M.~K.}\ \bibnamefont {Crawford}},\ }\href
  {http://dx.doi.org/10.1038/srep35685} {\bibfield  {journal} {\bibinfo
  {journal} {Sci. Rep.}\ }\textbf {\bibinfo {volume} {6}},\ \bibinfo {pages}
  {35685 EP } (\bibinfo {year} {2016})}\BibitemShut {NoStop}%
\bibitem [{\citenamefont {Gao}\ \emph {et~al.}(2016)\citenamefont {Gao},
  \citenamefont {Gao}, \citenamefont {Abtew}, \citenamefont {Sun},
  \citenamefont {Zhang},\ and\ \citenamefont {Zhang}}]{Gao}%
  \BibitemOpen
  \bibfield  {author} {\bibinfo {author} {\bibfnamefont {W.}~\bibnamefont
  {Gao}}, \bibinfo {author} {\bibfnamefont {X.}~\bibnamefont {Gao}}, \bibinfo
  {author} {\bibfnamefont {T.~A.}\ \bibnamefont {Abtew}}, \bibinfo {author}
  {\bibfnamefont {Y.-Y.}\ \bibnamefont {Sun}}, \bibinfo {author} {\bibfnamefont
  {S.}~\bibnamefont {Zhang}}, \ and\ \bibinfo {author} {\bibfnamefont
  {P.}~\bibnamefont {Zhang}},\ }\href {\doibase 10.1103/PhysRevB.93.085202}
  {\bibfield  {journal} {\bibinfo  {journal} {Phys. Rev. B}\ }\textbf {\bibinfo
  {volume} {93}},\ \bibinfo {pages} {085202} (\bibinfo {year}
  {2016})}\BibitemShut {NoStop}%
\bibitem [{\citenamefont {Zhu}\ \emph {et~al.}(2016)\citenamefont {Zhu},
  \citenamefont {Miyata}, \citenamefont {Fu}, \citenamefont {Wang},
  \citenamefont {Joshi}, \citenamefont {Niesner}, \citenamefont {Williams},
  \citenamefont {Jin},\ and\ \citenamefont {Zhu}}]{Zhu}%
  \BibitemOpen
  \bibfield  {author} {\bibinfo {author} {\bibfnamefont {H.}~\bibnamefont
  {Zhu}}, \bibinfo {author} {\bibfnamefont {K.}~\bibnamefont {Miyata}},
  \bibinfo {author} {\bibfnamefont {Y.}~\bibnamefont {Fu}}, \bibinfo {author}
  {\bibfnamefont {J.}~\bibnamefont {Wang}}, \bibinfo {author} {\bibfnamefont
  {P.~P.}\ \bibnamefont {Joshi}}, \bibinfo {author} {\bibfnamefont
  {D.}~\bibnamefont {Niesner}}, \bibinfo {author} {\bibfnamefont {K.~W.}\
  \bibnamefont {Williams}}, \bibinfo {author} {\bibfnamefont {S.}~\bibnamefont
  {Jin}}, \ and\ \bibinfo {author} {\bibfnamefont {X.-Y.}\ \bibnamefont
  {Zhu}},\ }\href {\doibase 10.1126/science.aaf9570} {\bibfield  {journal}
  {\bibinfo  {journal} {Science}\ }\textbf {\bibinfo {volume} {353}},\ \bibinfo
  {pages} {1409} (\bibinfo {year} {2016})}\BibitemShut {NoStop}%
\bibitem [{\citenamefont {Fujii}\ \emph {et~al.}(1974)\citenamefont {Fujii},
  \citenamefont {Hoshino}, \citenamefont {Yamada},\ and\ \citenamefont
  {Shirane}}]{Fuji}%
  \BibitemOpen
  \bibfield  {author} {\bibinfo {author} {\bibfnamefont {Y.}~\bibnamefont
  {Fujii}}, \bibinfo {author} {\bibfnamefont {S.}~\bibnamefont {Hoshino}},
  \bibinfo {author} {\bibfnamefont {Y.}~\bibnamefont {Yamada}}, \ and\ \bibinfo
  {author} {\bibfnamefont {G.}~\bibnamefont {Shirane}},\ }\href {\doibase
  10.1103/PhysRevB.9.4549} {\bibfield  {journal} {\bibinfo  {journal} {Phys.
  Rev. B}\ }\textbf {\bibinfo {volume} {9}},\ \bibinfo {pages} {4549} (\bibinfo
  {year} {1974})}\BibitemShut {NoStop}%
\bibitem [{\citenamefont {Brivio}\ \emph {et~al.}(2015)\citenamefont {Brivio},
  \citenamefont {Frost}, \citenamefont {Skelton}, \citenamefont {Jackson},
  \citenamefont {Weber}, \citenamefont {Weller}, \citenamefont {Go\~ni},
  \citenamefont {Leguy}, \citenamefont {Barnes},\ and\ \citenamefont
  {Walsh}}]{Brivio}%
  \BibitemOpen
  \bibfield  {author} {\bibinfo {author} {\bibfnamefont {F.}~\bibnamefont
  {Brivio}}, \bibinfo {author} {\bibfnamefont {J.~M.}\ \bibnamefont {Frost}},
  \bibinfo {author} {\bibfnamefont {J.~M.}\ \bibnamefont {Skelton}}, \bibinfo
  {author} {\bibfnamefont {A.~J.}\ \bibnamefont {Jackson}}, \bibinfo {author}
  {\bibfnamefont {O.~J.}\ \bibnamefont {Weber}}, \bibinfo {author}
  {\bibfnamefont {M.~T.}\ \bibnamefont {Weller}}, \bibinfo {author}
  {\bibfnamefont {A.~R.}\ \bibnamefont {Go\~ni}}, \bibinfo {author}
  {\bibfnamefont {A.~M.~A.}\ \bibnamefont {Leguy}}, \bibinfo {author}
  {\bibfnamefont {P.~R.~F.}\ \bibnamefont {Barnes}}, \ and\ \bibinfo {author}
  {\bibfnamefont {A.}~\bibnamefont {Walsh}},\ }\href {\doibase
  10.1103/PhysRevB.92.144308} {\bibfield  {journal} {\bibinfo  {journal} {Phys.
  Rev. B}\ }\textbf {\bibinfo {volume} {92}},\ \bibinfo {pages} {144308}
  (\bibinfo {year} {2015})}\BibitemShut {NoStop}%
\bibitem [{\citenamefont {Comin}\ \emph {et~al.}(2016)\citenamefont {Comin},
  \citenamefont {Crawford}, \citenamefont {Said}, \citenamefont {Herron},
  \citenamefont {Guise}, \citenamefont {Wang}, \citenamefont {Whitfield},
  \citenamefont {Jain}, \citenamefont {Gong}, \citenamefont {McGaughey},\ and\
  \citenamefont {Sargent}}]{Comin}%
  \BibitemOpen
  \bibfield  {author} {\bibinfo {author} {\bibfnamefont {R.}~\bibnamefont
  {Comin}}, \bibinfo {author} {\bibfnamefont {M.~K.}\ \bibnamefont {Crawford}},
  \bibinfo {author} {\bibfnamefont {A.~H.}\ \bibnamefont {Said}}, \bibinfo
  {author} {\bibfnamefont {N.}~\bibnamefont {Herron}}, \bibinfo {author}
  {\bibfnamefont {W.~E.}\ \bibnamefont {Guise}}, \bibinfo {author}
  {\bibfnamefont {X.}~\bibnamefont {Wang}}, \bibinfo {author} {\bibfnamefont
  {P.~S.}\ \bibnamefont {Whitfield}}, \bibinfo {author} {\bibfnamefont
  {A.}~\bibnamefont {Jain}}, \bibinfo {author} {\bibfnamefont {X.}~\bibnamefont
  {Gong}}, \bibinfo {author} {\bibfnamefont {A.~J.~H.}\ \bibnamefont
  {McGaughey}}, \ and\ \bibinfo {author} {\bibfnamefont {E.~H.}\ \bibnamefont
  {Sargent}},\ }\href {\doibase 10.1103/PhysRevB.94.094301} {\bibfield
  {journal} {\bibinfo  {journal} {Phys. Rev. B}\ }\textbf {\bibinfo {volume}
  {94}},\ \bibinfo {pages} {094301} (\bibinfo {year} {2016})}\BibitemShut
  {NoStop}%
\bibitem [{\citenamefont {Lee}\ \emph {et~al.}(2015)\citenamefont {Lee},
  \citenamefont {Bristowe}, \citenamefont {Bristowe},\ and\ \citenamefont
  {Cheetham}}]{Lee}%
  \BibitemOpen
  \bibfield  {author} {\bibinfo {author} {\bibfnamefont {J.-H.}\ \bibnamefont
  {Lee}}, \bibinfo {author} {\bibfnamefont {N.~C.}\ \bibnamefont {Bristowe}},
  \bibinfo {author} {\bibfnamefont {P.~D.}\ \bibnamefont {Bristowe}}, \ and\
  \bibinfo {author} {\bibfnamefont {A.~K.}\ \bibnamefont {Cheetham}},\ }\href
  {\doibase 10.1039/C5CC00979K} {\bibfield  {journal} {\bibinfo  {journal}
  {Chem. Commun.}\ }\textbf {\bibinfo {volume} {51}},\ \bibinfo {pages} {6434}
  (\bibinfo {year} {2015})}\BibitemShut {NoStop}%
\bibitem [{\citenamefont {Lee}\ \emph {et~al.}(2016)\citenamefont {Lee},
  \citenamefont {Bristowe}, \citenamefont {Lee}, \citenamefont {Lee},
  \citenamefont {Bristowe}, \citenamefont {Cheetham},\ and\ \citenamefont
  {Jang}}]{Lee2}%
  \BibitemOpen
  \bibfield  {author} {\bibinfo {author} {\bibfnamefont {J.-H.}\ \bibnamefont
  {Lee}}, \bibinfo {author} {\bibfnamefont {N.~C.}\ \bibnamefont {Bristowe}},
  \bibinfo {author} {\bibfnamefont {J.~H.}\ \bibnamefont {Lee}}, \bibinfo
  {author} {\bibfnamefont {S.-H.}\ \bibnamefont {Lee}}, \bibinfo {author}
  {\bibfnamefont {P.~D.}\ \bibnamefont {Bristowe}}, \bibinfo {author}
  {\bibfnamefont {A.~K.}\ \bibnamefont {Cheetham}}, \ and\ \bibinfo {author}
  {\bibfnamefont {H.~M.}\ \bibnamefont {Jang}},\ }\href {\doibase
  10.1021/acs.chemmater.6b00968} {\bibfield  {journal} {\bibinfo  {journal}
  {Chem. Mater.}\ }\textbf {\bibinfo {volume} {28}},\ \bibinfo {pages} {4259}
  (\bibinfo {year} {2016})}\BibitemShut {NoStop}%
\bibitem [{\citenamefont {L{\'e}toublon}\ \emph {et~al.}(2016)\citenamefont
  {L{\'e}toublon}, \citenamefont {Paofai}, \citenamefont {Ruffl{\'e}},
  \citenamefont {Bourges}, \citenamefont {Hehlen}, \citenamefont {Michel},
  \citenamefont {Ecolivet}, \citenamefont {Durand}, \citenamefont {Cordier},
  \citenamefont {Katan},\ and\ \citenamefont {Even}}]{Letoubon}%
  \BibitemOpen
  \bibfield  {author} {\bibinfo {author} {\bibfnamefont {A.}~\bibnamefont
  {L{\'e}toublon}}, \bibinfo {author} {\bibfnamefont {S.}~\bibnamefont
  {Paofai}}, \bibinfo {author} {\bibfnamefont {B.}~\bibnamefont {Ruffl{\'e}}},
  \bibinfo {author} {\bibfnamefont {P.}~\bibnamefont {Bourges}}, \bibinfo
  {author} {\bibfnamefont {B.}~\bibnamefont {Hehlen}}, \bibinfo {author}
  {\bibfnamefont {T.}~\bibnamefont {Michel}}, \bibinfo {author} {\bibfnamefont
  {C.}~\bibnamefont {Ecolivet}}, \bibinfo {author} {\bibfnamefont
  {O.}~\bibnamefont {Durand}}, \bibinfo {author} {\bibfnamefont
  {S.}~\bibnamefont {Cordier}}, \bibinfo {author} {\bibfnamefont
  {C.}~\bibnamefont {Katan}}, \ and\ \bibinfo {author} {\bibfnamefont
  {J.}~\bibnamefont {Even}},\ }\href {\doibase 10.1021/acs.jpclett.6b01709}
  {\bibfield  {journal} {\bibinfo  {journal} {J. Phys. Chem. Lett.}\ }\textbf
  {\bibinfo {volume} {7}},\ \bibinfo {pages} {3776} (\bibinfo {year}
  {2016})}\BibitemShut {NoStop}%
\bibitem [{\citenamefont {Swainson}\ \emph {et~al.}(2015)\citenamefont
  {Swainson}, \citenamefont {Stock}, \citenamefont {Parker}, \citenamefont
  {Van~Eijck}, \citenamefont {Russina},\ and\ \citenamefont
  {Taylor}}]{Swainson}%
  \BibitemOpen
  \bibfield  {author} {\bibinfo {author} {\bibfnamefont {I.~P.}\ \bibnamefont
  {Swainson}}, \bibinfo {author} {\bibfnamefont {C.}~\bibnamefont {Stock}},
  \bibinfo {author} {\bibfnamefont {S.~F.}\ \bibnamefont {Parker}}, \bibinfo
  {author} {\bibfnamefont {L.}~\bibnamefont {Van~Eijck}}, \bibinfo {author}
  {\bibfnamefont {M.}~\bibnamefont {Russina}}, \ and\ \bibinfo {author}
  {\bibfnamefont {J.~W.}\ \bibnamefont {Taylor}},\ }\href {\doibase
  10.1103/PhysRevB.92.100303} {\bibfield  {journal} {\bibinfo  {journal} {Phys.
  Rev. B}\ }\textbf {\bibinfo {volume} {92}},\ \bibinfo {pages} {100303}
  (\bibinfo {year} {2015})}\BibitemShut {NoStop}%
\bibitem [{\citenamefont {Kawamura}\ and\ \citenamefont
  {Mashiyama}(1999)}]{Kawamura}%
  \BibitemOpen
  \bibfield  {author} {\bibinfo {author} {\bibfnamefont {Y.}~\bibnamefont
  {Kawamura}}\ and\ \bibinfo {author} {\bibfnamefont {H.}~\bibnamefont
  {Mashiyama}},\ }\href
  {https://www.researchgate.net/publication/290695747_Modulated_structure_in_phase_II_of_CH3NH3PbCl3}
  {\bibfield  {journal} {\bibinfo  {journal} {J. Korean Phys. Soc.}\ }\textbf
  {\bibinfo {volume} {35}},\ \bibinfo {pages} {S1437} (\bibinfo {year}
  {1999})}\BibitemShut {NoStop}%
\bibitem [{\citenamefont {Ong}\ \emph {et~al.}(2015)\citenamefont {Ong},
  \citenamefont {Goh}, \citenamefont {Xu},\ and\ \citenamefont {Huan}}]{Ong}%
  \BibitemOpen
  \bibfield  {author} {\bibinfo {author} {\bibfnamefont {K.~P.}\ \bibnamefont
  {Ong}}, \bibinfo {author} {\bibfnamefont {T.~W.}\ \bibnamefont {Goh}},
  \bibinfo {author} {\bibfnamefont {Q.}~\bibnamefont {Xu}}, \ and\ \bibinfo
  {author} {\bibfnamefont {A.}~\bibnamefont {Huan}},\ }\href {\doibase
  10.1021/jz502740d} {\bibfield  {journal} {\bibinfo  {journal} {J. Phys. Chem.
  Lett.}\ }\textbf {\bibinfo {volume} {6}},\ \bibinfo {pages} {681} (\bibinfo
  {year} {2015})}\BibitemShut {NoStop}%
\bibitem [{\citenamefont {Mashiyama}, \citenamefont {Kawamura},\ and\
  \citenamefont {Kubota}(2007)}]{Mashiyama2}%
  \BibitemOpen
  \bibfield  {author} {\bibinfo {author} {\bibfnamefont {H.}~\bibnamefont
  {Mashiyama}}, \bibinfo {author} {\bibfnamefont {Y.}~\bibnamefont {Kawamura}},
  \ and\ \bibinfo {author} {\bibfnamefont {Y.}~\bibnamefont {Kubota}},\ }\href
  {\doibase 10.3938/jkps.51.850} {\bibfield  {journal} {\bibinfo  {journal} {J.
  Korean Phys. Soc.}\ }\textbf {\bibinfo {volume} {51}},\ \bibinfo {pages}
  {850} (\bibinfo {year} {2007})}\BibitemShut {NoStop}%
\bibitem [{\citenamefont {Page}\ \emph {et~al.}(2016)\citenamefont {Page},
  \citenamefont {Siewenie}, \citenamefont {Quadrelli},\ and\ \citenamefont
  {Malavasi}}]{Page}%
  \BibitemOpen
  \bibfield  {author} {\bibinfo {author} {\bibfnamefont {K.}~\bibnamefont
  {Page}}, \bibinfo {author} {\bibfnamefont {J.~E.}\ \bibnamefont {Siewenie}},
  \bibinfo {author} {\bibfnamefont {P.}~\bibnamefont {Quadrelli}}, \ and\
  \bibinfo {author} {\bibfnamefont {L.}~\bibnamefont {Malavasi}},\ }\href
  {\doibase 10.1002/anie.201608602} {\bibfield  {journal} {\bibinfo  {journal}
  {Angew. Chem. Int. Ed.}\ }\textbf {\bibinfo {volume} {55}},\ \bibinfo {pages}
  {14320} (\bibinfo {year} {2016})}\BibitemShut {NoStop}%
\bibitem [{\citenamefont {Swainson}\ \emph {et~al.}(2003)\citenamefont
  {Swainson}, \citenamefont {Hammond}, \citenamefont {Soulli{\`e}re},
  \citenamefont {Knop},\ and\ \citenamefont {Massa}}]{Swainson2}%
  \BibitemOpen
  \bibfield  {author} {\bibinfo {author} {\bibfnamefont {I.}~\bibnamefont
  {Swainson}}, \bibinfo {author} {\bibfnamefont {R.}~\bibnamefont {Hammond}},
  \bibinfo {author} {\bibfnamefont {C.}~\bibnamefont {Soulli{\`e}re}}, \bibinfo
  {author} {\bibfnamefont {O.}~\bibnamefont {Knop}}, \ and\ \bibinfo {author}
  {\bibfnamefont {W.}~\bibnamefont {Massa}},\ }\href {\doibase
  https://doi.org/10.1016/S0022-4596(03)00352-9} {\bibfield  {journal}
  {\bibinfo  {journal} {J. Solid State Chem.}\ }\textbf {\bibinfo {volume}
  {176}},\ \bibinfo {pages} {97 } (\bibinfo {year} {2003})}\BibitemShut
  {NoStop}%
\bibitem [{\citenamefont {Brock}\ \emph
  {et~al.}(1989{\natexlab{b}})\citenamefont {Brock}, \citenamefont {Birgeneau},
  \citenamefont {Litster},\ and\ \citenamefont {Aharony}}]{Brock2}%
  \BibitemOpen
  \bibfield  {author} {\bibinfo {author} {\bibfnamefont {J.~D.}\ \bibnamefont
  {Brock}}, \bibinfo {author} {\bibfnamefont {R.~J.}\ \bibnamefont
  {Birgeneau}}, \bibinfo {author} {\bibfnamefont {D.}~\bibnamefont {Litster}},
  \ and\ \bibinfo {author} {\bibfnamefont {A.}~\bibnamefont {Aharony}},\ }\href
  {\doibase 10.1080/00107518908213772} {\bibfield  {journal} {\bibinfo
  {journal} {Contemp. Phys.}\ }\textbf {\bibinfo {volume} {30}},\ \bibinfo
  {pages} {321} (\bibinfo {year} {1989}{\natexlab{b}})}\BibitemShut {NoStop}%
\bibitem [{\citenamefont {Als-Nielsen}\ \emph {et~al.}(1980)\citenamefont
  {Als-Nielsen}, \citenamefont {Litster}, \citenamefont {Birgeneau},
  \citenamefont {Kaplan}, \citenamefont {Safinya}, \citenamefont
  {Lindegaard-Andersen},\ and\ \citenamefont {Mathiesen}}]{AlsNielsen}%
  \BibitemOpen
  \bibfield  {author} {\bibinfo {author} {\bibfnamefont {J.}~\bibnamefont
  {Als-Nielsen}}, \bibinfo {author} {\bibfnamefont {J.~D.}\ \bibnamefont
  {Litster}}, \bibinfo {author} {\bibfnamefont {R.~J.}\ \bibnamefont
  {Birgeneau}}, \bibinfo {author} {\bibfnamefont {M.}~\bibnamefont {Kaplan}},
  \bibinfo {author} {\bibfnamefont {C.~R.}\ \bibnamefont {Safinya}}, \bibinfo
  {author} {\bibfnamefont {A.}~\bibnamefont {Lindegaard-Andersen}}, \ and\
  \bibinfo {author} {\bibfnamefont {S.}~\bibnamefont {Mathiesen}},\ }\href
  {\doibase 10.1103/PhysRevB.22.312} {\bibfield  {journal} {\bibinfo  {journal}
  {Phys. Rev. B}\ }\textbf {\bibinfo {volume} {22}},\ \bibinfo {pages} {312}
  (\bibinfo {year} {1980})}\BibitemShut {NoStop}%
\bibitem [{\citenamefont {Pindak}\ \emph {et~al.}(1980)\citenamefont {Pindak},
  \citenamefont {Young}, \citenamefont {Meyer},\ and\ \citenamefont
  {Clark}}]{Pindak}%
  \BibitemOpen
  \bibfield  {author} {\bibinfo {author} {\bibfnamefont {R.}~\bibnamefont
  {Pindak}}, \bibinfo {author} {\bibfnamefont {C.~Y.}\ \bibnamefont {Young}},
  \bibinfo {author} {\bibfnamefont {R.~B.}\ \bibnamefont {Meyer}}, \ and\
  \bibinfo {author} {\bibfnamefont {N.~A.}\ \bibnamefont {Clark}},\ }\href
  {\doibase 10.1103/PhysRevLett.45.1193} {\bibfield  {journal} {\bibinfo
  {journal} {Phys. Rev. Lett.}\ }\textbf {\bibinfo {volume} {45}},\ \bibinfo
  {pages} {1193} (\bibinfo {year} {1980})}\BibitemShut {NoStop}%
\bibitem [{\citenamefont {Hagen}\ \emph {et~al.}(1992)\citenamefont {Hagen},
  \citenamefont {Dove}, \citenamefont {Harris}, \citenamefont {Steigenberger},\
  and\ \citenamefont {Powell}}]{Hagen}%
  \BibitemOpen
  \bibfield  {author} {\bibinfo {author} {\bibfnamefont {M.}~\bibnamefont
  {Hagen}}, \bibinfo {author} {\bibfnamefont {M.}~\bibnamefont {Dove}},
  \bibinfo {author} {\bibfnamefont {M.}~\bibnamefont {Harris}}, \bibinfo
  {author} {\bibfnamefont {U.}~\bibnamefont {Steigenberger}}, \ and\ \bibinfo
  {author} {\bibfnamefont {B.}~\bibnamefont {Powell}},\ }\href {\doibase
  http://dx.doi.org/10.1016/0921-4526(92)90732-8} {\bibfield  {journal}
  {\bibinfo  {journal} {Physica B}\ }\textbf {\bibinfo {volume} {180}},\
  \bibinfo {pages} {276 } (\bibinfo {year} {1992})}\BibitemShut {NoStop}%
\bibitem [{\citenamefont {Harris}\ \emph {et~al.}(1998)\citenamefont {Harris},
  \citenamefont {Dove}, \citenamefont {Swainson},\ and\ \citenamefont
  {Hagen}}]{Harris}%
  \BibitemOpen
  \bibfield  {author} {\bibinfo {author} {\bibfnamefont {M.~J.}\ \bibnamefont
  {Harris}}, \bibinfo {author} {\bibfnamefont {M.~T.}\ \bibnamefont {Dove}},
  \bibinfo {author} {\bibfnamefont {I.~P.}\ \bibnamefont {Swainson}}, \ and\
  \bibinfo {author} {\bibfnamefont {M.~E.}\ \bibnamefont {Hagen}},\ }\href
  {http://stacks.iop.org/0953-8984/10/i=25/a=002} {\bibfield  {journal}
  {\bibinfo  {journal} {J. Phys. Condens. Matter}\ }\textbf {\bibinfo {volume}
  {10}},\ \bibinfo {pages} {L423} (\bibinfo {year} {1998})}\BibitemShut
  {NoStop}%
\bibitem [{\citenamefont {Shirane}\ \emph {et~al.}(1970)\citenamefont
  {Shirane}, \citenamefont {Axe}, \citenamefont {Harada},\ and\ \citenamefont
  {Remeika}}]{Shirane}%
  \BibitemOpen
  \bibfield  {author} {\bibinfo {author} {\bibfnamefont {G.}~\bibnamefont
  {Shirane}}, \bibinfo {author} {\bibfnamefont {J.~D.}\ \bibnamefont {Axe}},
  \bibinfo {author} {\bibfnamefont {J.}~\bibnamefont {Harada}}, \ and\ \bibinfo
  {author} {\bibfnamefont {J.~P.}\ \bibnamefont {Remeika}},\ }\href {\doibase
  10.1103/PhysRevB.2.155} {\bibfield  {journal} {\bibinfo  {journal} {Phys.
  Rev. B}\ }\textbf {\bibinfo {volume} {2}},\ \bibinfo {pages} {155} (\bibinfo
  {year} {1970})}\BibitemShut {NoStop}%
\bibitem [{\citenamefont {Leguy}\ \emph {et~al.}(2016)\citenamefont {Leguy},
  \citenamefont {Goni}, \citenamefont {Frost}, \citenamefont {Skelton},
  \citenamefont {Brivio}, \citenamefont {Rodriguez-Martinez}, \citenamefont
  {Weber}, \citenamefont {Pallipurath}, \citenamefont {Alonso}, \citenamefont
  {Campoy-Quiles}, \citenamefont {Weller}, \citenamefont {Nelson},
  \citenamefont {Walsh},\ and\ \citenamefont {Barnes}}]{Leguy}%
  \BibitemOpen
  \bibfield  {author} {\bibinfo {author} {\bibfnamefont {A.~M.~A.}\
  \bibnamefont {Leguy}}, \bibinfo {author} {\bibfnamefont {A.~R.}\ \bibnamefont
  {Goni}}, \bibinfo {author} {\bibfnamefont {J.~M.}\ \bibnamefont {Frost}},
  \bibinfo {author} {\bibfnamefont {J.}~\bibnamefont {Skelton}}, \bibinfo
  {author} {\bibfnamefont {F.}~\bibnamefont {Brivio}}, \bibinfo {author}
  {\bibfnamefont {X.}~\bibnamefont {Rodriguez-Martinez}}, \bibinfo {author}
  {\bibfnamefont {O.~J.}\ \bibnamefont {Weber}}, \bibinfo {author}
  {\bibfnamefont {A.}~\bibnamefont {Pallipurath}}, \bibinfo {author}
  {\bibfnamefont {M.~I.}\ \bibnamefont {Alonso}}, \bibinfo {author}
  {\bibfnamefont {M.}~\bibnamefont {Campoy-Quiles}}, \bibinfo {author}
  {\bibfnamefont {M.~T.}\ \bibnamefont {Weller}}, \bibinfo {author}
  {\bibfnamefont {J.}~\bibnamefont {Nelson}}, \bibinfo {author} {\bibfnamefont
  {A.}~\bibnamefont {Walsh}}, \ and\ \bibinfo {author} {\bibfnamefont
  {P.~R.~F.}\ \bibnamefont {Barnes}},\ }\href {\doibase 10.1039/C6CP03474H}
  {\bibfield  {journal} {\bibinfo  {journal} {Phys. Chem. Chem. Phys.}\
  }\textbf {\bibinfo {volume} {18}},\ \bibinfo {pages} {27051} (\bibinfo {year}
  {2016})}\BibitemShut {NoStop}%
\bibitem [{\citenamefont {Dru{\.z}bicki}\ \emph {et~al.}(2016)\citenamefont
  {Dru{\.z}bicki}, \citenamefont {Pinna}, \citenamefont {Rudi{\'c}},
  \citenamefont {Jura}, \citenamefont {Gorini},\ and\ \citenamefont
  {Fernandez-Alonso}}]{Druzbiki}%
  \BibitemOpen
  \bibfield  {author} {\bibinfo {author} {\bibfnamefont {K.}~\bibnamefont
  {Dru{\.z}bicki}}, \bibinfo {author} {\bibfnamefont {R.~S.}\ \bibnamefont
  {Pinna}}, \bibinfo {author} {\bibfnamefont {S.}~\bibnamefont {Rudi{\'c}}},
  \bibinfo {author} {\bibfnamefont {M.}~\bibnamefont {Jura}}, \bibinfo {author}
  {\bibfnamefont {G.}~\bibnamefont {Gorini}}, \ and\ \bibinfo {author}
  {\bibfnamefont {F.}~\bibnamefont {Fernandez-Alonso}},\ }\href {\doibase
  10.1021/acs.jpclett.6b01822} {\bibfield  {journal} {\bibinfo  {journal} {J.
  Phys. Chem. Lett.}\ }\textbf {\bibinfo {volume} {7}},\ \bibinfo {pages}
  {4701} (\bibinfo {year} {2016})}\BibitemShut {NoStop}%
\bibitem [{\citenamefont {Campbell}, \citenamefont {Telling},\ and\
  \citenamefont {Carlile}(2000)}]{Campbell}%
  \BibitemOpen
  \bibfield  {author} {\bibinfo {author} {\bibfnamefont {S.}~\bibnamefont
  {Campbell}}, \bibinfo {author} {\bibfnamefont {M.}~\bibnamefont {Telling}}, \
  and\ \bibinfo {author} {\bibfnamefont {C.}~\bibnamefont {Carlile}},\ }\href
  {http://www.sciencedirect.com/science/article/pii/S0921452699012867}
  {\bibfield  {journal} {\bibinfo  {journal} {Physica B}\ }\textbf {\bibinfo
  {volume} {276}},\ \bibinfo {pages} {206 } (\bibinfo {year}
  {2000})}\BibitemShut {NoStop}%
\bibitem [{\citenamefont {Perring}\ \emph {et~al.}(1994)\citenamefont
  {Perring}, \citenamefont {Taylor}, \citenamefont {McK~Paul}, \citenamefont
  {Boothroyd},\ and\ \citenamefont {Aeppli}}]{Perring}%
  \BibitemOpen
  \bibfield  {author} {\bibinfo {author} {\bibfnamefont {T.~G.}\ \bibnamefont
  {Perring}}, \bibinfo {author} {\bibfnamefont {A.~D.}\ \bibnamefont {Taylor}},
  \bibinfo {author} {\bibfnamefont {D.}~\bibnamefont {McK~Paul}}, \bibinfo
  {author} {\bibfnamefont {A.~T.}\ \bibnamefont {Boothroyd}}, \ and\ \bibinfo
  {author} {\bibfnamefont {G.}~\bibnamefont {Aeppli}},\ }\href
  {http://neutronresearch.com/parch/1993/01/199301010600.pdf} {\bibfield
  {journal} {\bibinfo  {journal} {ICANS XII}\ ,\ \bibinfo {pages} {I}}
  (\bibinfo {year} {1994})}\BibitemShut {NoStop}%
\bibitem [{\citenamefont {Adams}\ \emph {et~al.}(2009)\citenamefont {Adams},
  \citenamefont {Parker}, \citenamefont {Fernandez-Alonso}, \citenamefont
  {Cutler}, \citenamefont {Hodges},\ and\ \citenamefont {King}}]{Adams}%
  \BibitemOpen
  \bibfield  {author} {\bibinfo {author} {\bibfnamefont {M.~A.}\ \bibnamefont
  {Adams}}, \bibinfo {author} {\bibfnamefont {S.~F.}\ \bibnamefont {Parker}},
  \bibinfo {author} {\bibfnamefont {F.}~\bibnamefont {Fernandez-Alonso}},
  \bibinfo {author} {\bibfnamefont {D.~J.}\ \bibnamefont {Cutler}}, \bibinfo
  {author} {\bibfnamefont {C.}~\bibnamefont {Hodges}}, \ and\ \bibinfo {author}
  {\bibfnamefont {A.}~\bibnamefont {King}},\ }\href
  {http://as.osa.org/abstract.cfm?URI=as-63-7-727} {\bibfield  {journal}
  {\bibinfo  {journal} {Appl. Spectrosc.}\ }\textbf {\bibinfo {volume} {63}},\
  \bibinfo {pages} {727} (\bibinfo {year} {2009})}\BibitemShut {NoStop}%
\bibitem [{\citenamefont {Sears}(1992)}]{Sears}%
  \BibitemOpen
  \bibfield  {author} {\bibinfo {author} {\bibfnamefont {V.~F.}\ \bibnamefont
  {Sears}},\ }\href {\doibase 10.1080/10448639208218770} {\bibfield  {journal}
  {\bibinfo  {journal} {Neutron News}\ }\textbf {\bibinfo {volume} {3}},\
  \bibinfo {pages} {26} (\bibinfo {year} {1992})}\BibitemShut {NoStop}%
\bibitem [{\citenamefont {Bakulin}\ \emph {et~al.}(2015)\citenamefont
  {Bakulin}, \citenamefont {Selig}, \citenamefont {Bakker}, \citenamefont
  {Rezus}, \citenamefont {M{\"u}ller}, \citenamefont {Glaser}, \citenamefont
  {Lovrincic}, \citenamefont {Sun}, \citenamefont {Chen}, \citenamefont
  {Walsh}, \citenamefont {Frost},\ and\ \citenamefont {Jansen}}]{Bakulin}%
  \BibitemOpen
  \bibfield  {author} {\bibinfo {author} {\bibfnamefont {A.~A.}\ \bibnamefont
  {Bakulin}}, \bibinfo {author} {\bibfnamefont {O.}~\bibnamefont {Selig}},
  \bibinfo {author} {\bibfnamefont {H.~J.}\ \bibnamefont {Bakker}}, \bibinfo
  {author} {\bibfnamefont {Y.~L.}\ \bibnamefont {Rezus}}, \bibinfo {author}
  {\bibfnamefont {C.}~\bibnamefont {M{\"u}ller}}, \bibinfo {author}
  {\bibfnamefont {T.}~\bibnamefont {Glaser}}, \bibinfo {author} {\bibfnamefont
  {R.}~\bibnamefont {Lovrincic}}, \bibinfo {author} {\bibfnamefont
  {Z.}~\bibnamefont {Sun}}, \bibinfo {author} {\bibfnamefont {Z.}~\bibnamefont
  {Chen}}, \bibinfo {author} {\bibfnamefont {A.}~\bibnamefont {Walsh}},
  \bibinfo {author} {\bibfnamefont {J.~M.}\ \bibnamefont {Frost}}, \ and\
  \bibinfo {author} {\bibfnamefont {T.~L.~C.}\ \bibnamefont {Jansen}},\ }\href
  {\doibase 10.1021/acs.jpclett.5b01555} {\bibfield  {journal} {\bibinfo
  {journal} {The Journal of Physical Chemistry Letters}\ }\textbf {\bibinfo
  {volume} {6}},\ \bibinfo {pages} {3663} (\bibinfo {year} {2015})}\BibitemShut
  {NoStop}%
\bibitem [{\citenamefont {Knop}\ \emph {et~al.}(1990)\citenamefont {Knop},
  \citenamefont {Wasylishen}, \citenamefont {White}, \citenamefont {Cameron},\
  and\ \citenamefont {Oort}}]{Knop}%
  \BibitemOpen
  \bibfield  {author} {\bibinfo {author} {\bibfnamefont {O.}~\bibnamefont
  {Knop}}, \bibinfo {author} {\bibfnamefont {R.~E.}\ \bibnamefont
  {Wasylishen}}, \bibinfo {author} {\bibfnamefont {M.~A.}\ \bibnamefont
  {White}}, \bibinfo {author} {\bibfnamefont {T.~S.}\ \bibnamefont {Cameron}},
  \ and\ \bibinfo {author} {\bibfnamefont {M.~J. M.~V.}\ \bibnamefont {Oort}},\
  }\href {\doibase 10.1139/v90-063} {\bibfield  {journal} {\bibinfo  {journal}
  {Canadian Journal of Chemistry}\ }\textbf {\bibinfo {volume} {68}},\ \bibinfo
  {pages} {412} (\bibinfo {year} {1990})}\BibitemShut {NoStop}%
\bibitem [{\citenamefont {Leguy}\ \emph {et~al.}(2015)\citenamefont {Leguy},
  \citenamefont {Frost}, \citenamefont {McMahon}, \citenamefont {Sakai},
  \citenamefont {Kockelmann}, \citenamefont {Law}, \citenamefont {Li},
  \citenamefont {Foglia}, \citenamefont {Walsh}, \citenamefont {O'Regan},
  \citenamefont {Nelson}, \citenamefont {Cabral},\ and\ \citenamefont
  {Barnes}}]{Leguy2}%
  \BibitemOpen
  \bibfield  {author} {\bibinfo {author} {\bibfnamefont {A.~M.~A.}\
  \bibnamefont {Leguy}}, \bibinfo {author} {\bibfnamefont {J.~M.}\ \bibnamefont
  {Frost}}, \bibinfo {author} {\bibfnamefont {A.~P.}\ \bibnamefont {McMahon}},
  \bibinfo {author} {\bibfnamefont {V.~G.}\ \bibnamefont {Sakai}}, \bibinfo
  {author} {\bibfnamefont {W.}~\bibnamefont {Kockelmann}}, \bibinfo {author}
  {\bibfnamefont {C.}~\bibnamefont {Law}}, \bibinfo {author} {\bibfnamefont
  {X.}~\bibnamefont {Li}}, \bibinfo {author} {\bibfnamefont {F.}~\bibnamefont
  {Foglia}}, \bibinfo {author} {\bibfnamefont {A.}~\bibnamefont {Walsh}},
  \bibinfo {author} {\bibfnamefont {B.~C.}\ \bibnamefont {O'Regan}}, \bibinfo
  {author} {\bibfnamefont {J.}~\bibnamefont {Nelson}}, \bibinfo {author}
  {\bibfnamefont {J.~T.}\ \bibnamefont {Cabral}}, \ and\ \bibinfo {author}
  {\bibfnamefont {P.~R.~F.}\ \bibnamefont {Barnes}},\ }\href
  {http://dx.doi.org/10.1038/ncomms8124} {\bibfield  {journal} {\bibinfo
  {journal} {Nat. Commun.}\ }\textbf {\bibinfo {volume} {6}},\ \bibinfo {pages}
  {7124 EP } (\bibinfo {year} {2015})}\BibitemShut {NoStop}%
\bibitem [{\citenamefont {Weller}\ \emph {et~al.}(2015)\citenamefont {Weller},
  \citenamefont {Weber}, \citenamefont {Henry}, \citenamefont {Di~Pumpo},\ and\
  \citenamefont {Hansen}}]{Weller}%
  \BibitemOpen
  \bibfield  {author} {\bibinfo {author} {\bibfnamefont {M.~T.}\ \bibnamefont
  {Weller}}, \bibinfo {author} {\bibfnamefont {O.~J.}\ \bibnamefont {Weber}},
  \bibinfo {author} {\bibfnamefont {P.~F.}\ \bibnamefont {Henry}}, \bibinfo
  {author} {\bibfnamefont {A.~M.}\ \bibnamefont {Di~Pumpo}}, \ and\ \bibinfo
  {author} {\bibfnamefont {T.~C.}\ \bibnamefont {Hansen}},\ }\href {\doibase
  10.1039/C4CC09944C} {\bibfield  {journal} {\bibinfo  {journal} {Chem.
  Commun.}\ }\textbf {\bibinfo {volume} {51}},\ \bibinfo {pages} {4180}
  (\bibinfo {year} {2015})}\BibitemShut {NoStop}%
\bibitem [{\citenamefont {Chen}\ \emph {et~al.}(2015)\citenamefont {Chen},
  \citenamefont {Foley}, \citenamefont {Ipek}, \citenamefont {Tyagi},
  \citenamefont {Copley}, \citenamefont {Brown}, \citenamefont {Choi},\ and\
  \citenamefont {Lee}}]{Chen}%
  \BibitemOpen
  \bibfield  {author} {\bibinfo {author} {\bibfnamefont {T.}~\bibnamefont
  {Chen}}, \bibinfo {author} {\bibfnamefont {B.~J.}\ \bibnamefont {Foley}},
  \bibinfo {author} {\bibfnamefont {B.}~\bibnamefont {Ipek}}, \bibinfo {author}
  {\bibfnamefont {M.}~\bibnamefont {Tyagi}}, \bibinfo {author} {\bibfnamefont
  {J.~R.~D.}\ \bibnamefont {Copley}}, \bibinfo {author} {\bibfnamefont {C.~M.}\
  \bibnamefont {Brown}}, \bibinfo {author} {\bibfnamefont {J.~J.}\ \bibnamefont
  {Choi}}, \ and\ \bibinfo {author} {\bibfnamefont {S.-H.}\ \bibnamefont
  {Lee}},\ }\href {\doibase 10.1039/C5CP05348J} {\bibfield  {journal} {\bibinfo
   {journal} {Phys. Chem. Chem. Phys.}\ }\textbf {\bibinfo {volume} {17}},\
  \bibinfo {pages} {31278} (\bibinfo {year} {2015})}\BibitemShut {NoStop}%
\bibitem [{\citenamefont {Maaej}\ \emph {et~al.}(1998)\citenamefont {Maaej},
  \citenamefont {Bahri}, \citenamefont {Abid}, \citenamefont {Jaidane},
  \citenamefont {Lakhdar},\ and\ \citenamefont {Lauti{\'e}}}]{Maaej}%
  \BibitemOpen
  \bibfield  {author} {\bibinfo {author} {\bibfnamefont {A.}~\bibnamefont
  {Maaej}}, \bibinfo {author} {\bibfnamefont {M.}~\bibnamefont {Bahri}},
  \bibinfo {author} {\bibfnamefont {Y.}~\bibnamefont {Abid}}, \bibinfo {author}
  {\bibfnamefont {N.}~\bibnamefont {Jaidane}}, \bibinfo {author} {\bibfnamefont
  {Z.~B.}\ \bibnamefont {Lakhdar}}, \ and\ \bibinfo {author} {\bibfnamefont
  {A.}~\bibnamefont {Lauti{\'e}}},\ }\href {\doibase 10.1080/01411599808207997}
  {\bibfield  {journal} {\bibinfo  {journal} {Phase Transit.}\ }\textbf
  {\bibinfo {volume} {64}},\ \bibinfo {pages} {179} (\bibinfo {year}
  {1998})}\BibitemShut {NoStop}%
\bibitem [{\citenamefont {Pistor}\ \emph {et~al.}(2016)\citenamefont {Pistor},
  \citenamefont {Ruiz}, \citenamefont {Cabot},\ and\ \citenamefont
  {Izquierdo-Roca}}]{Pistor}%
  \BibitemOpen
  \bibfield  {author} {\bibinfo {author} {\bibfnamefont {P.}~\bibnamefont
  {Pistor}}, \bibinfo {author} {\bibfnamefont {A.}~\bibnamefont {Ruiz}},
  \bibinfo {author} {\bibfnamefont {A.}~\bibnamefont {Cabot}}, \ and\ \bibinfo
  {author} {\bibfnamefont {V.}~\bibnamefont {Izquierdo-Roca}},\ }\href
  {http://dx.doi.org/10.1038/srep35973} {\bibfield  {journal} {\bibinfo
  {journal} {Sci. Rep.}\ }\textbf {\bibinfo {volume} {6}},\ \bibinfo {pages}
  {35973 EP } (\bibinfo {year} {2016})}\BibitemShut {NoStop}%
\bibitem [{\citenamefont {Yaffe}\ \emph {et~al.}(2017)\citenamefont {Yaffe},
  \citenamefont {Guo}, \citenamefont {Tan}, \citenamefont {Egger},
  \citenamefont {Hull}, \citenamefont {Stoumpos}, \citenamefont {Zheng},
  \citenamefont {Heinz}, \citenamefont {Kronik}, \citenamefont {Kanatzidis},
  \citenamefont {Owen}, \citenamefont {Rappe}, \citenamefont {Pimenta},\ and\
  \citenamefont {Brus}}]{Yaffe}%
  \BibitemOpen
  \bibfield  {author} {\bibinfo {author} {\bibfnamefont {O.}~\bibnamefont
  {Yaffe}}, \bibinfo {author} {\bibfnamefont {Y.}~\bibnamefont {Guo}}, \bibinfo
  {author} {\bibfnamefont {L.~Z.}\ \bibnamefont {Tan}}, \bibinfo {author}
  {\bibfnamefont {D.~A.}\ \bibnamefont {Egger}}, \bibinfo {author}
  {\bibfnamefont {T.}~\bibnamefont {Hull}}, \bibinfo {author} {\bibfnamefont
  {C.~C.}\ \bibnamefont {Stoumpos}}, \bibinfo {author} {\bibfnamefont
  {F.}~\bibnamefont {Zheng}}, \bibinfo {author} {\bibfnamefont {T.~F.}\
  \bibnamefont {Heinz}}, \bibinfo {author} {\bibfnamefont {L.}~\bibnamefont
  {Kronik}}, \bibinfo {author} {\bibfnamefont {M.~G.}\ \bibnamefont
  {Kanatzidis}}, \bibinfo {author} {\bibfnamefont {J.~S.}\ \bibnamefont
  {Owen}}, \bibinfo {author} {\bibfnamefont {A.~M.}\ \bibnamefont {Rappe}},
  \bibinfo {author} {\bibfnamefont {M.~A.}\ \bibnamefont {Pimenta}}, \ and\
  \bibinfo {author} {\bibfnamefont {L.~E.}\ \bibnamefont {Brus}},\ }\href
  {\doibase 10.1103/PhysRevLett.118.136001} {\bibfield  {journal} {\bibinfo
  {journal} {Phys. Rev. Lett.}\ }\textbf {\bibinfo {volume} {118}},\ \bibinfo
  {pages} {136001} (\bibinfo {year} {2017})}\BibitemShut {NoStop}%
\bibitem [{\citenamefont {{Guo}}\ \emph {et~al.}(2017)\citenamefont {{Guo}},
  \citenamefont {{Yaffe}}, \citenamefont {{Paley}}, \citenamefont {{Beecher}},
  \citenamefont {{Hull}}, \citenamefont {{Szpak}}, \citenamefont {{Owen}},
  \citenamefont {{Brus}},\ and\ \citenamefont {{Pimenta}}}]{Guo}%
  \BibitemOpen
  \bibfield  {author} {\bibinfo {author} {\bibfnamefont {Y.}~\bibnamefont
  {{Guo}}}, \bibinfo {author} {\bibfnamefont {O.}~\bibnamefont {{Yaffe}}},
  \bibinfo {author} {\bibfnamefont {D.~W.}\ \bibnamefont {{Paley}}}, \bibinfo
  {author} {\bibfnamefont {A.~N.}\ \bibnamefont {{Beecher}}}, \bibinfo {author}
  {\bibfnamefont {T.~D.}\ \bibnamefont {{Hull}}}, \bibinfo {author}
  {\bibfnamefont {G.}~\bibnamefont {{Szpak}}}, \bibinfo {author} {\bibfnamefont
  {J.~S.}\ \bibnamefont {{Owen}}}, \bibinfo {author} {\bibfnamefont {L.~E.}\
  \bibnamefont {{Brus}}}, \ and\ \bibinfo {author} {\bibfnamefont {M.~A.}\
  \bibnamefont {{Pimenta}}},\ }\href
  {http://adsabs.harvard.edu/abs/2017arXiv170510691G} {\bibfield  {journal}
  {\bibinfo  {journal} {ArXiv}\ } (\bibinfo {year} {2017})}\BibitemShut
  {NoStop}%
\bibitem [{\citenamefont {Stock}\ \emph {et~al.}(2010)\citenamefont {Stock},
  \citenamefont {Cowley}, \citenamefont {Taylor},\ and\ \citenamefont
  {Bennington}}]{Stock}%
  \BibitemOpen
  \bibfield  {author} {\bibinfo {author} {\bibfnamefont {C.}~\bibnamefont
  {Stock}}, \bibinfo {author} {\bibfnamefont {R.~A.}\ \bibnamefont {Cowley}},
  \bibinfo {author} {\bibfnamefont {J.~W.}\ \bibnamefont {Taylor}}, \ and\
  \bibinfo {author} {\bibfnamefont {S.~M.}\ \bibnamefont {Bennington}},\ }\href
  {\doibase 10.1103/PhysRevB.81.024303} {\bibfield  {journal} {\bibinfo
  {journal} {Phys. Rev. B}\ }\textbf {\bibinfo {volume} {81}},\ \bibinfo
  {pages} {024303} (\bibinfo {year} {2010})}\BibitemShut {NoStop}%
\bibitem [{\citenamefont {Niemann}\ \emph {et~al.}(2016)\citenamefont
  {Niemann}, \citenamefont {Kontos}, \citenamefont {Palles}, \citenamefont
  {Kamitsos}, \citenamefont {Kaltzoglou}, \citenamefont {Brivio}, \citenamefont
  {Falaras},\ and\ \citenamefont {Cameron}}]{Niemann}%
  \BibitemOpen
  \bibfield  {author} {\bibinfo {author} {\bibfnamefont {R.~G.}\ \bibnamefont
  {Niemann}}, \bibinfo {author} {\bibfnamefont {A.~G.}\ \bibnamefont {Kontos}},
  \bibinfo {author} {\bibfnamefont {D.}~\bibnamefont {Palles}}, \bibinfo
  {author} {\bibfnamefont {E.~I.}\ \bibnamefont {Kamitsos}}, \bibinfo {author}
  {\bibfnamefont {A.}~\bibnamefont {Kaltzoglou}}, \bibinfo {author}
  {\bibfnamefont {F.}~\bibnamefont {Brivio}}, \bibinfo {author} {\bibfnamefont
  {P.}~\bibnamefont {Falaras}}, \ and\ \bibinfo {author} {\bibfnamefont
  {P.~J.}\ \bibnamefont {Cameron}},\ }\href {\doibase 10.1021/acs.jpcc.5b11256}
  {\bibfield  {journal} {\bibinfo  {journal} {The Journal of Physical Chemistry
  C}\ }\textbf {\bibinfo {volume} {120}},\ \bibinfo {pages} {2509} (\bibinfo
  {year} {2016})}\BibitemShut {NoStop}%
\bibitem [{\citenamefont {Chen}\ \emph {et~al.}(2017)\citenamefont {Chen},
  \citenamefont {Mo}, \citenamefont {Yang}, \citenamefont {Zhou},\ and\
  \citenamefont {Gao}}]{Chen2}%
  \BibitemOpen
  \bibfield  {author} {\bibinfo {author} {\bibfnamefont {J.}~\bibnamefont
  {Chen}}, \bibinfo {author} {\bibfnamefont {Z.-H.}\ \bibnamefont {Mo}},
  \bibinfo {author} {\bibfnamefont {X.}~\bibnamefont {Yang}}, \bibinfo {author}
  {\bibfnamefont {H.-L.}\ \bibnamefont {Zhou}}, \ and\ \bibinfo {author}
  {\bibfnamefont {Q.}~\bibnamefont {Gao}},\ }\href {\doibase
  10.1039/C7CC02782F} {\bibfield  {journal} {\bibinfo  {journal} {Chem.
  Commun.}\ }\textbf {\bibinfo {volume} {53}},\ \bibinfo {pages} {6949}
  (\bibinfo {year} {2017})}\BibitemShut {NoStop}%
\bibitem [{\citenamefont {Quan}\ \emph {et~al.}(2016)\citenamefont {Quan},
  \citenamefont {Yuan}, \citenamefont {Comin}, \citenamefont {Voznyy},
  \citenamefont {Beauregard}, \citenamefont {Hoogland}, \citenamefont {Buin},
  \citenamefont {Kirmani}, \citenamefont {Zhao}, \citenamefont {Amassian},
  \citenamefont {Kim},\ and\ \citenamefont {Sargent}}]{Quan}%
  \BibitemOpen
  \bibfield  {author} {\bibinfo {author} {\bibfnamefont {L.~N.}\ \bibnamefont
  {Quan}}, \bibinfo {author} {\bibfnamefont {M.}~\bibnamefont {Yuan}}, \bibinfo
  {author} {\bibfnamefont {R.}~\bibnamefont {Comin}}, \bibinfo {author}
  {\bibfnamefont {O.}~\bibnamefont {Voznyy}}, \bibinfo {author} {\bibfnamefont
  {E.~M.}\ \bibnamefont {Beauregard}}, \bibinfo {author} {\bibfnamefont
  {S.}~\bibnamefont {Hoogland}}, \bibinfo {author} {\bibfnamefont
  {A.}~\bibnamefont {Buin}}, \bibinfo {author} {\bibfnamefont {A.~R.}\
  \bibnamefont {Kirmani}}, \bibinfo {author} {\bibfnamefont {K.}~\bibnamefont
  {Zhao}}, \bibinfo {author} {\bibfnamefont {A.}~\bibnamefont {Amassian}},
  \bibinfo {author} {\bibfnamefont {D.~H.}\ \bibnamefont {Kim}}, \ and\
  \bibinfo {author} {\bibfnamefont {E.~H.}\ \bibnamefont {Sargent}},\ }\href
  {\doibase 10.1021/jacs.5b11740} {\bibfield  {journal} {\bibinfo  {journal}
  {J. Am. Chem. Soc.}\ }\textbf {\bibinfo {volume} {138}},\ \bibinfo {pages}
  {2649} (\bibinfo {year} {2016})}\BibitemShut {NoStop}%
\bibitem [{\citenamefont {Roiland}\ \emph {et~al.}(2016)\citenamefont
  {Roiland}, \citenamefont {Trippe-Allard}, \citenamefont {Jemli},
  \citenamefont {Alonso}, \citenamefont {Ameline}, \citenamefont {Gautier},
  \citenamefont {Bataille}, \citenamefont {Le~Polles}, \citenamefont
  {Deleporte}, \citenamefont {Even},\ and\ \citenamefont {Katan}}]{Roiland}%
  \BibitemOpen
  \bibfield  {author} {\bibinfo {author} {\bibfnamefont {C.}~\bibnamefont
  {Roiland}}, \bibinfo {author} {\bibfnamefont {G.}~\bibnamefont
  {Trippe-Allard}}, \bibinfo {author} {\bibfnamefont {K.}~\bibnamefont
  {Jemli}}, \bibinfo {author} {\bibfnamefont {B.}~\bibnamefont {Alonso}},
  \bibinfo {author} {\bibfnamefont {J.-C.}\ \bibnamefont {Ameline}}, \bibinfo
  {author} {\bibfnamefont {R.}~\bibnamefont {Gautier}}, \bibinfo {author}
  {\bibfnamefont {T.}~\bibnamefont {Bataille}}, \bibinfo {author}
  {\bibfnamefont {L.}~\bibnamefont {Le~Polles}}, \bibinfo {author}
  {\bibfnamefont {E.}~\bibnamefont {Deleporte}}, \bibinfo {author}
  {\bibfnamefont {J.}~\bibnamefont {Even}}, \ and\ \bibinfo {author}
  {\bibfnamefont {C.}~\bibnamefont {Katan}},\ }\href {\doibase
  10.1039/C6CP02947G} {\bibfield  {journal} {\bibinfo  {journal} {Phys. Chem.
  Chem. Phys.}\ }\textbf {\bibinfo {volume} {18}},\ \bibinfo {pages} {27133}
  (\bibinfo {year} {2016})}\BibitemShut {NoStop}%
\bibitem [{\citenamefont {Mattoni}\ \emph {et~al.}(2015)\citenamefont
  {Mattoni}, \citenamefont {Filippetti}, \citenamefont {Saba},\ and\
  \citenamefont {Delugas}}]{Mattoni}%
  \BibitemOpen
  \bibfield  {author} {\bibinfo {author} {\bibfnamefont {A.}~\bibnamefont
  {Mattoni}}, \bibinfo {author} {\bibfnamefont {A.}~\bibnamefont {Filippetti}},
  \bibinfo {author} {\bibfnamefont {M.~I.}\ \bibnamefont {Saba}}, \ and\
  \bibinfo {author} {\bibfnamefont {P.}~\bibnamefont {Delugas}},\ }\href
  {\doibase 10.1021/acs.jpcc.5b04283} {\bibfield  {journal} {\bibinfo
  {journal} {J. Phys. Chem. C}\ }\textbf {\bibinfo {volume} {119}},\ \bibinfo
  {pages} {17421} (\bibinfo {year} {2015})}\BibitemShut {NoStop}%
\bibitem [{\citenamefont {Motta}\ \emph {et~al.}(2015)\citenamefont {Motta},
  \citenamefont {El-Mellouhi}, \citenamefont {Kais}, \citenamefont {Tabet},
  \citenamefont {Alharbi},\ and\ \citenamefont {Sanvito}}]{Motta}%
  \BibitemOpen
  \bibfield  {author} {\bibinfo {author} {\bibfnamefont {C.}~\bibnamefont
  {Motta}}, \bibinfo {author} {\bibfnamefont {F.}~\bibnamefont {El-Mellouhi}},
  \bibinfo {author} {\bibfnamefont {S.}~\bibnamefont {Kais}}, \bibinfo {author}
  {\bibfnamefont {N.}~\bibnamefont {Tabet}}, \bibinfo {author} {\bibfnamefont
  {F.}~\bibnamefont {Alharbi}}, \ and\ \bibinfo {author} {\bibfnamefont
  {S.}~\bibnamefont {Sanvito}},\ }\href {http://dx.doi.org/10.1038/ncomms8026}
  {\bibfield  {journal} {\bibinfo  {journal} {Nat. Commun.}\ }\textbf {\bibinfo
  {volume} {6}},\ \bibinfo {pages} {7026 EP } (\bibinfo {year}
  {2015})}\BibitemShut {NoStop}%
\bibitem [{\citenamefont {Even}, \citenamefont {Carignano},\ and\ \citenamefont
  {Katan}(2016)}]{Even}%
  \BibitemOpen
  \bibfield  {author} {\bibinfo {author} {\bibfnamefont {J.}~\bibnamefont
  {Even}}, \bibinfo {author} {\bibfnamefont {M.}~\bibnamefont {Carignano}}, \
  and\ \bibinfo {author} {\bibfnamefont {C.}~\bibnamefont {Katan}},\ }\href
  {\doibase 10.1039/C5NR06386H} {\bibfield  {journal} {\bibinfo  {journal}
  {Nanoscale}\ }\textbf {\bibinfo {volume} {8}},\ \bibinfo {pages} {6222}
  (\bibinfo {year} {2016})}\BibitemShut {NoStop}%
\bibitem [{\citenamefont {Niesner}\ \emph {et~al.}(2016)\citenamefont
  {Niesner}, \citenamefont {Wilhelm}, \citenamefont {Levchuk}, \citenamefont
  {Osvet}, \citenamefont {Shrestha}, \citenamefont {Batentschuk}, \citenamefont
  {Brabec},\ and\ \citenamefont {Fauster}}]{Niesner}%
  \BibitemOpen
  \bibfield  {author} {\bibinfo {author} {\bibfnamefont {D.}~\bibnamefont
  {Niesner}}, \bibinfo {author} {\bibfnamefont {M.}~\bibnamefont {Wilhelm}},
  \bibinfo {author} {\bibfnamefont {I.}~\bibnamefont {Levchuk}}, \bibinfo
  {author} {\bibfnamefont {A.}~\bibnamefont {Osvet}}, \bibinfo {author}
  {\bibfnamefont {S.}~\bibnamefont {Shrestha}}, \bibinfo {author}
  {\bibfnamefont {M.}~\bibnamefont {Batentschuk}}, \bibinfo {author}
  {\bibfnamefont {C.}~\bibnamefont {Brabec}}, \ and\ \bibinfo {author}
  {\bibfnamefont {T.}~\bibnamefont {Fauster}},\ }\href {\doibase
  10.1103/PhysRevLett.117.126401} {\bibfield  {journal} {\bibinfo  {journal}
  {Phys. Rev. Lett.}\ }\textbf {\bibinfo {volume} {117}},\ \bibinfo {pages}
  {126401} (\bibinfo {year} {2016})}\BibitemShut {NoStop}%
\bibitem [{\citenamefont {Even}\ \emph {et~al.}(2016)\citenamefont {Even},
  \citenamefont {Paofai}, \citenamefont {Bourges}, \citenamefont {Letoublon},
  \citenamefont {Cordier}, \citenamefont {Durand},\ and\ \citenamefont
  {Katan}}]{Even2}%
  \BibitemOpen
  \bibfield  {author} {\bibinfo {author} {\bibfnamefont {J.}~\bibnamefont
  {Even}}, \bibinfo {author} {\bibfnamefont {S.}~\bibnamefont {Paofai}},
  \bibinfo {author} {\bibfnamefont {P.}~\bibnamefont {Bourges}}, \bibinfo
  {author} {\bibfnamefont {A.}~\bibnamefont {Letoublon}}, \bibinfo {author}
  {\bibfnamefont {S.}~\bibnamefont {Cordier}}, \bibinfo {author} {\bibfnamefont
  {O.}~\bibnamefont {Durand}}, \ and\ \bibinfo {author} {\bibfnamefont
  {C.}~\bibnamefont {Katan}},\ }\href {\doibase 10.1117/12.2213623} {\bibfield
  {journal} {\bibinfo  {journal} {Proc. SPIE}\ }\textbf {\bibinfo {volume}
  {9743}},\ \bibinfo {pages} {97430M} (\bibinfo {year} {2016})}\BibitemShut
  {NoStop}%
\bibitem [{\citenamefont {Minns}\ \emph {et~al.}(2017)\citenamefont {Minns},
  \citenamefont {Zajdel}, \citenamefont {Chernyshov}, \citenamefont {van
  Beek},\ and\ \citenamefont {Green}}]{Minns}%
  \BibitemOpen
  \bibfield  {author} {\bibinfo {author} {\bibfnamefont {J.~L.}\ \bibnamefont
  {Minns}}, \bibinfo {author} {\bibfnamefont {P.}~\bibnamefont {Zajdel}},
  \bibinfo {author} {\bibfnamefont {D.}~\bibnamefont {Chernyshov}}, \bibinfo
  {author} {\bibfnamefont {W.}~\bibnamefont {van Beek}}, \ and\ \bibinfo
  {author} {\bibfnamefont {M.~A.}\ \bibnamefont {Green}},\ }\href
  {http://dx.doi.org/10.1038/ncomms15152} {\bibfield  {journal} {\bibinfo
  {journal} {Nat. Commun.}\ }\textbf {\bibinfo {volume} {8}},\ \bibinfo {pages}
  {15152 EP } (\bibinfo {year} {2017})}\BibitemShut {NoStop}%
\bibitem [{\citenamefont {Fabini}\ \emph {et~al.}(2016)\citenamefont {Fabini},
  \citenamefont {Hogan}, \citenamefont {Evans}, \citenamefont {Stoumpos},
  \citenamefont {Kanatzidis},\ and\ \citenamefont {Seshadri}}]{Fabini}%
  \BibitemOpen
  \bibfield  {author} {\bibinfo {author} {\bibfnamefont {D.~H.}\ \bibnamefont
  {Fabini}}, \bibinfo {author} {\bibfnamefont {T.}~\bibnamefont {Hogan}},
  \bibinfo {author} {\bibfnamefont {H.~A.}\ \bibnamefont {Evans}}, \bibinfo
  {author} {\bibfnamefont {C.~C.}\ \bibnamefont {Stoumpos}}, \bibinfo {author}
  {\bibfnamefont {M.~G.}\ \bibnamefont {Kanatzidis}}, \ and\ \bibinfo {author}
  {\bibfnamefont {R.}~\bibnamefont {Seshadri}},\ }\href {\doibase
  10.1021/acs.jpclett.5b02821} {\bibfield  {journal} {\bibinfo  {journal} {J.
  Phys. Chem. Lett.}\ }\textbf {\bibinfo {volume} {7}},\ \bibinfo {pages} {376}
  (\bibinfo {year} {2016})}\BibitemShut {NoStop}%
\bibitem [{\citenamefont {Lynden-Bell}\ and\ \citenamefont
  {Michel}(1994)}]{Bell}%
  \BibitemOpen
  \bibfield  {author} {\bibinfo {author} {\bibfnamefont {R.~M.}\ \bibnamefont
  {Lynden-Bell}}\ and\ \bibinfo {author} {\bibfnamefont {K.~H.}\ \bibnamefont
  {Michel}},\ }\href {\doibase 10.1103/RevModPhys.66.721} {\bibfield  {journal}
  {\bibinfo  {journal} {Rev. Mod. Phys.}\ }\textbf {\bibinfo {volume} {66}},\
  \bibinfo {pages} {721} (\bibinfo {year} {1994})}\BibitemShut {NoStop}%
\bibitem [{\citenamefont {Endres}\ \emph {et~al.}(2016)\citenamefont {Endres},
  \citenamefont {Egger}, \citenamefont {Kulbak}, \citenamefont {Kerner},
  \citenamefont {Zhao}, \citenamefont {Silver}, \citenamefont {Hodes},
  \citenamefont {Rand}, \citenamefont {Cahen}, \citenamefont {Kronik},\ and\
  \citenamefont {Kahn}}]{Endres}%
  \BibitemOpen
  \bibfield  {author} {\bibinfo {author} {\bibfnamefont {J.}~\bibnamefont
  {Endres}}, \bibinfo {author} {\bibfnamefont {D.~A.}\ \bibnamefont {Egger}},
  \bibinfo {author} {\bibfnamefont {M.}~\bibnamefont {Kulbak}}, \bibinfo
  {author} {\bibfnamefont {R.~A.}\ \bibnamefont {Kerner}}, \bibinfo {author}
  {\bibfnamefont {L.}~\bibnamefont {Zhao}}, \bibinfo {author} {\bibfnamefont
  {S.~H.}\ \bibnamefont {Silver}}, \bibinfo {author} {\bibfnamefont
  {G.}~\bibnamefont {Hodes}}, \bibinfo {author} {\bibfnamefont {B.~P.}\
  \bibnamefont {Rand}}, \bibinfo {author} {\bibfnamefont {D.}~\bibnamefont
  {Cahen}}, \bibinfo {author} {\bibfnamefont {L.}~\bibnamefont {Kronik}}, \
  and\ \bibinfo {author} {\bibfnamefont {A.}~\bibnamefont {Kahn}},\ }\href
  {\doibase 10.1021/acs.jpclett.6b00946} {\bibfield  {journal} {\bibinfo
  {journal} {J. Phys. Chem. Lett.}\ }\textbf {\bibinfo {volume} {7}},\ \bibinfo
  {pages} {2722} (\bibinfo {year} {2016})}\BibitemShut {NoStop}%
\bibitem [{\citenamefont {Wright}\ \emph {et~al.}(2016)\citenamefont {Wright},
  \citenamefont {Verdi}, \citenamefont {Milot}, \citenamefont {Eperon},
  \citenamefont {P{\'e}rez-Osorio}, \citenamefont {Snaith}, \citenamefont
  {Giustino}, \citenamefont {Johnston},\ and\ \citenamefont {Herz}}]{Wright}%
  \BibitemOpen
  \bibfield  {author} {\bibinfo {author} {\bibfnamefont {A.~D.}\ \bibnamefont
  {Wright}}, \bibinfo {author} {\bibfnamefont {C.}~\bibnamefont {Verdi}},
  \bibinfo {author} {\bibfnamefont {R.~L.}\ \bibnamefont {Milot}}, \bibinfo
  {author} {\bibfnamefont {G.~E.}\ \bibnamefont {Eperon}}, \bibinfo {author}
  {\bibfnamefont {M.~A.}\ \bibnamefont {P{\'e}rez-Osorio}}, \bibinfo {author}
  {\bibfnamefont {H.~J.}\ \bibnamefont {Snaith}}, \bibinfo {author}
  {\bibfnamefont {F.}~\bibnamefont {Giustino}}, \bibinfo {author}
  {\bibfnamefont {M.~B.}\ \bibnamefont {Johnston}}, \ and\ \bibinfo {author}
  {\bibfnamefont {L.~M.}\ \bibnamefont {Herz}},\ }\href
  {http://dx.doi.org/10.1038/ncomms11755} {\bibfield  {journal} {\bibinfo
  {journal} {Nat. Commun.}\ }\textbf {\bibinfo {volume} {7}},\ \bibinfo {pages}
  {11755 EP } (\bibinfo {year} {2016})}\BibitemShut {NoStop}%
\bibitem [{\citenamefont {Butler}\ \emph {et~al.}(2016)\citenamefont {Butler},
  \citenamefont {Svane}, \citenamefont {Kieslich}, \citenamefont {Cheetham},\
  and\ \citenamefont {Walsh}}]{Butler}%
  \BibitemOpen
  \bibfield  {author} {\bibinfo {author} {\bibfnamefont {K.~T.}\ \bibnamefont
  {Butler}}, \bibinfo {author} {\bibfnamefont {K.}~\bibnamefont {Svane}},
  \bibinfo {author} {\bibfnamefont {G.}~\bibnamefont {Kieslich}}, \bibinfo
  {author} {\bibfnamefont {A.~K.}\ \bibnamefont {Cheetham}}, \ and\ \bibinfo
  {author} {\bibfnamefont {A.}~\bibnamefont {Walsh}},\ }\href {\doibase
  10.1103/PhysRevB.94.180103} {\bibfield  {journal} {\bibinfo  {journal} {Phys.
  Rev. B}\ }\textbf {\bibinfo {volume} {94}},\ \bibinfo {pages} {180103}
  (\bibinfo {year} {2016})}\BibitemShut {NoStop}%
\end{thebibliography}

%

\end{document}